\documentclass[%
 reprint,
 amsmath,amssymb,
 aps,
]{revtex4-2}

\usepackage{graphicx}
\usepackage{dcolumn}
\usepackage{bm}
\usepackage{color}
\usepackage{xcolor}
\usepackage[normalem]{ulem}

\newcommand{\Planck}{{\it Planck}}
\newcommand{\Keck}{{\it{Keck}}}

\begin{document}


\title{Planck and BICEP/Keck Array 2018 constraints on primordial gravitational waves and perspectives for future B-mode polarization measurements}
\author{Daniela Paoletti}
\email{daniela.paoletti@inaf.it}
\affiliation{INAF/OAS Bologna, Osservatorio di Astrofisica e Scienza dello Spazio, Area della ricerca CNR-INAF, via Gobetti 101, I-40129 Bologna, Italy}
 \affiliation{INFN, Sezione di Bologna, Via Irnerio 46, I-40126, Bologna, Italy}
\author{Fabio Finelli}%
 \email{fabio.finelli@inaf.it}
 \affiliation{INAF/OAS Bologna, Osservatorio di Astrofisica e Scienza dello Spazio, Area della ricerca CNR-INAF, via Gobetti 101, I-40129 Bologna, Italy}
 \affiliation{INFN, Sezione di Bologna, Via Irnerio 46, I-40126, Bologna, Italy}
\author{Jussi Valiviita}
\email{jussi.valiviita@helsinki.fi}
\affiliation{Helsinki Institute of Physics, Gustaf Hällströmin katu 2, University of Helsinki, Helsinki, Finland} 
\author{Masashi Hazumi}
\email{masashi.hazumi@kek.jp}
\affiliation{International Center for Quantum-field Measurement Systems for Studies of the Universe and Particles (QUP), High Energy Accelerator Research Organization (KEK), Tsukuba, Ibaraki 305-0801, Japan}
\affiliation{Institute of Particle and Nuclear Studies (IPNS), High Energy Accelerator Research Organization (KEK), Tsukuba, Ibaraki 305-0801, Japan}
\affiliation{Japan Aerospace Exploration Agency (JAXA), Institute of Space and Astronautical Science (ISAS), Sagamihara, Kanagawa 252-5210, Japan}
\affiliation{Kavli Institute for the Physics and Mathematics of the Universe (Kavli IPMU, WPI), UTIAS, The University of Tokyo, Kashiwa, Chiba 277-8583, Japan}
\affiliation{The Graduate University for Advanced Studies (SOKENDAI), Miura District, Kanagawa 240-0115, Hayama, Japan}
\date{October 31, 2022}

\begin{abstract}
Current and future $B$-mode polarization data are the most powerful observables to  constrain gravitational waves from the early Universe.
We set conservative constraints on tensor modes when relaxing the inflationary consistency condition $n_{\rm t}=-r/8$ between the tensor tilt $n_{\rm t}$ and the tensor-to-scalar ratio $r$.
By adding a power-law spectrum of tensor perturbations to $\Lambda$CDM 
and parametrizing this tensor contribution by two independent primordial tensor-to-scalar ratios $(r_1,r_2)$ at 
$k_1 = 0.005\,$Mpc$^{-1}$ and $k_2 = 0.02\,$Mpc$^{-1}$,
\Planck\ and BICEP/\Keck\ Array 2018 data (BK18) lead to constraints $r_{0.005} < 0.030$ and $r_{0.02} < 0.098$ at 95\% confidence level. The corresponding upper bound $r_{0.01} < 0.039$ is by a factor of $2$ tighter than the one obtained with \Planck\ 2018 and the older BK15 data.
We then study the perspectives for future CMB experiments that will measure both the reionization bump and recombination peak of the $B$-mode polarization angular power spectrum, such as LiteBIRD. We test the robustness of the results to the choice of the scales for $(r_1,r_2)$ in these future perspectives. Whereas distinguishing $n_{\rm t}=-r/8$ from exact scale invariance is impossible as expected, we show
how radical, theoretically motivated departures from $n_{\rm t}=-r/8$, which are consistent with the current data, could be distinguished with LiteBIRD.
Moreover, LiteBIRD will be able to shrink the allowed parameter space area in the $(r_{0.005},r_{0.02})$ plane to 
less than one hundredth of the currently allowed area by \Planck\ 2018 and BK18.
\end{abstract}

\maketitle

\section{Introduction}

Primordial gravitational waves generated during inflation \cite{Starobinsky:1979ty} have a characteristic shape in the Cosmic Microwave Background 
(CMB) angular power spectra of temperature and polarization anisotropies which distinguishes them from scalar curvature perturbations. 
On top of temperature ($T$) and $E$-mode ($E$) polarization (also produced by curvature perturbations), 
the distinctive imprint of primordial gravitational waves is $B$-mode polarization \cite{Kamionkowski:1996zd,Seljak:1996gy}. 
\begin{figure*}
\centering
\includegraphics[width=0.85\textwidth]{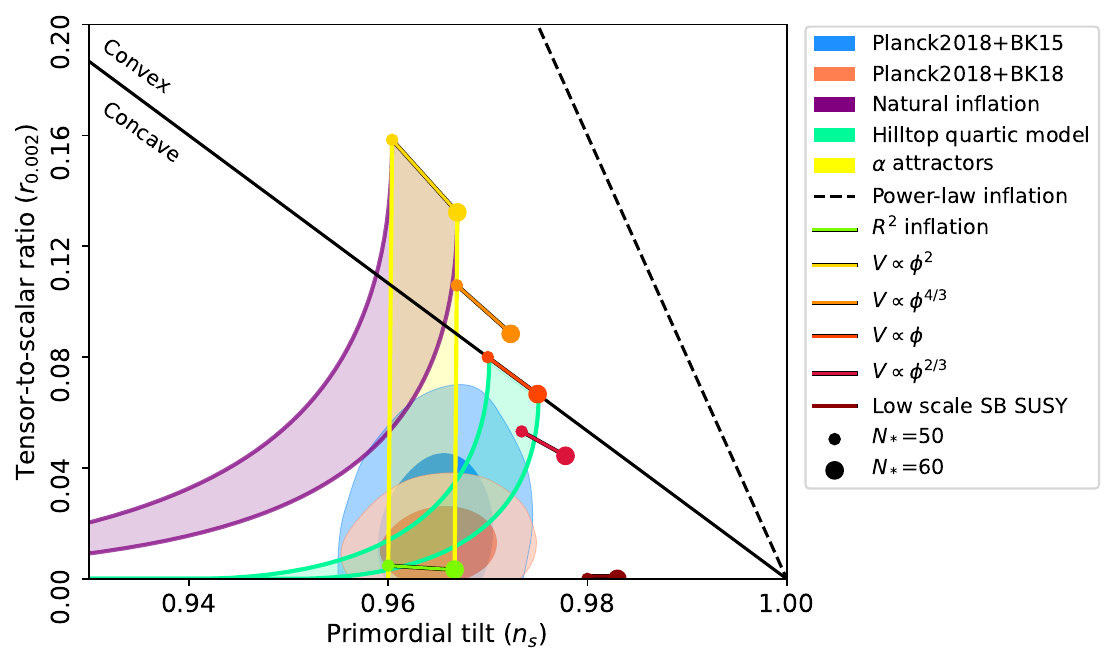}
\caption{Marginalized joint 68\% and 95\% CL regions in the $n_{\mathrm t} = -r/8$ model for $n_{\rm s}$ and $r_{0.002}$ from \Planck\ in combination with BK18 and BK15 data compared to the theoretical predictions of selected inflationary models with the uncertainty in the number of $e$-folds $N_*$ in the range $(50,\, 60)$.
}
\label{fig:nsr}
\end{figure*}
The increasing sensitivity of CMB polarization measurements has led to comparable constraints on primordial 
gravitational waves from $T$,$E$, and from $B$ mode separately by the joint analysis of BICEP2/\Keck\ Array and \Planck\ data \cite{BICEP2:2015nss}.
Since then, $B$-mode polarization alone
has given a tighter constraint than $T$,$E$ for the tensor-to-scalar ratio \cite{BICEP2:2015xme,BICEP2:2018kqh,BICEP:2021xfz},
\begin{equation}
    r(k) = \frac{{\mathcal P}_{\rm t}(k)}{{\mathcal P}_{\mathcal R}(k)}\,,
\label{ratio}
\end{equation}
where ${\mathcal P}_{\rm t}(k)$ and ${\mathcal P}_{\mathcal R}(k)$ are the tensor and scalar power spectra, here assumed as power laws,
\begin{equation}
 {\mathcal P}_{\rm t}(k) = A_{{\rm t}} \left( \frac{k}{k_{{\rm *}}}\right)^{n_{\rm t}}\,,
    \label{eq:PTk}
\end{equation}
\begin{equation}
    {\mathcal P}_{\mathcal R}(k) = A_{\mathcal R} \left( \frac{k}{k_{{\rm *}}}\right)^{n_{\rm s}-1}\,. 
    \label{eq:PSk}
\end{equation}
In the above, $n_{\rm t}$ ($n_{\rm s}$) is the tensor (scalar) spectral index and $A_{{\rm t}}$ ($A_{{\rm s}}$) the tensor (scalar) amplitude at the pivot scale, typically chosen such that $k_*=0.05\,$Mpc$^{-1}$
\footnote{We denote the tensor-to-scalar ratio at this scale simply by $r$ or occasionally by $r_{0.05}$ and at any other scale by adding a subscript indicating the corresponding wavenumber. Integer subscripts $1$ and $2$ refer to the scales of our two-scale parametrization, explained in Sec.~\ref{sec:two-scale}.}.

The recent release of BICEP-Keck Array data (BK18) \cite{BICEP:2021xfz} in combination with \Planck\  2018 data has set a 95\% CL upper limit on the tensor-to-scalar ratio $r < 0.035$ in the case of a scale-invariant primordial spectrum of gravitational waves. When the tensor tilt $n_{\mathrm t}$ satisfies the so-called consistency condition, i.e.,  
$n_{\mathrm t} = -r/8$, 
motivated by Bunch-Davies initial conditions for
tensor modes during slow-roll inflation driven by a single real scalar field with a standard
kinetic term (denoted in the following by SSSRI), the limit is unchanged.
This limit leads to the 95\% CL upper bound on the scale of inflation 
\begin{equation}
V_* = \frac{3 \pi^2 A_{\mathrm{s}}}{2} \, r \, M_{\mathrm {Pl}}^4 < (1.4 \times 10^{16}~{\mathrm{GeV}} )^4 \quad  (95\%\ \text{CL}),
\end{equation}
or on the Hubble parameter during inflation 
\begin{equation}
\frac{H_*}{M_{\mathrm{Pl}}} < 2.0 \times 10^{-5} \quad (95\%\ \text{CL}).
\end{equation}

The improvements with BK18 compared to BK15, when combined with \Planck\ 2018 data \cite{Planck:2018jri}, in terms of tighter constraints to slow-roll inflationary models can be seen in the $(n_{\mathrm{s}},\,r_{0.002})$ plane in Fig.~\ref{fig:nsr}.

The availability of an accurate and precise $B$-mode polarization likelihood has also made it possible to derive data driven constraints when the theoretical prior $n_{\mathrm t} = -r/8$ is relaxed \cite{Planck:2015sxf,Planck:2018jri}.
This more conservative and phenomenological approach is justified since deviations from $n_{\rm t}=-r/8$ are predicted in well-motivated theoretical inflationary models. These deviations occur, for example, with a non-standard kinetic term for a single scalar field \citep{Garriga:1999vw} or with a more general Lagrangian \citep{Kobayashi:2010cm}, with an initial vacuum state which is not Bunch-Davies \cite{Ashoorioon:2013eia}, when more than one scalar field is present \citep{Bartolo:2001rt,Wands:2002bn,Byrnes:2006fr} or these are coupled also through the kinetic terms \cite{DiMarco:2005nq,Achucarro:2012yr}, and in Gauge-flation when a non-Abelian gauge field in a particular isotropic configuration drives the accelerated stage \cite{Maleknejad:2011jw}. The relation $n_{\mathrm t} = -r/8$ is also violated when gravitational waves are not only amplified by the expansion from quantum fluctuations, but also sourced by spectator fields \citep{Cook:2011hg,Dimastrogiovanni:2016fuu,Agrawal:2017awz} present during inflation, an effect which also leads to significant primordial non-Gaussianity.
More radical departures from a nearly scale-invariant power spectrum are predicted in alternatives to inflation \citep{Gasperini:1992em,Boyle:2003km,Brandenberger:2006xi}.

In this paper we use the two-scale analysis for tensor perturbations \cite{Planck:2015sxf,CORE:2016ymi,Planck:2018jri} 
in order to present the updated BK18 conservative constraints and the perspectives for future CMB polarization measurements when the theoretical  prior $n_{\mathrm t} = -r/8$ is relaxed. 
After a review of the two-scale analysis for tensors in Sec.~\ref{sec:two-scale},  we present the \Planck\ 2018 + BK18 results in Sec.~\ref{sec:realdata} with a comparison to those derived with BK15 in \cite{Planck:2018jri}.
In Sec.~\ref{sec:forecasts}, we forecast the capability of future $B$-mode polarization measurements to constrain a power-law spectrum of gravitational waves, by taking as a representative example the specifications of the LiteBIRD mission \citep{LiteBIRD:2022cnt}.
We asses the dependence on the scales chosen and test how much the constraints would degrade if the low-multipole $B$-mode data
were missing.
In Sec.~\ref{sec:conclusions}, we draw conclusions. 

\begin{figure}
\includegraphics[width=0.48\textwidth]{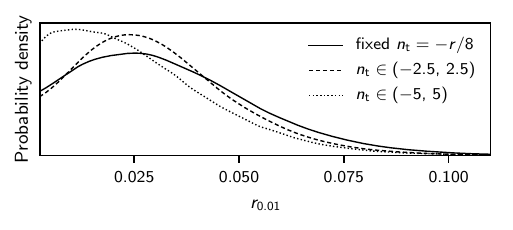}
\caption{A demonstration of the volume effect of the prior range of the tensor tilt 
on the tensor-to-scalar ratio when the
$(r,n_\mathrm{t})$ parametrization is adopted. 
In order to minimize other effects in this illustrative figure with the BK15 data, we have fixed other cosmological parameters to their \Planck\ 2018 best-fit values. We show the marginalized posterior pdf of $r_{0.01}$ from three MCMC runs: fixed $n_\mathrm{t}=-r/8$, a flat prior $(-2.5,\,2.5)$, or $(-5,\,5)$. 
The 95\% CL upper bound on $r_{0.01}$ is $0.073$, $0.065$, or $0.061$, respectively. (Note that a fixed $n_\mathrm{t}=0$ leads to a 95\% CL upper bound and a posterior pdf that are indistinguishable from the $n_\mathrm{t}=-r/8$ case.)\label{fig:ntprioreffect}}
\end{figure}

\section{Beyond consistency condition on the tensor tilt\label{sec:two-scale}}

In this paper, we relax the condition $n_\mathrm{t} = -r/8$ and let the data (either the real or simulated one) determine both the amplitude of tensor perturbations and the tilt of their spectrum $n_\mathrm{t}$.
These analyses are highly motivated by two reasons: 1) Without testing for this possibility, one would not know if there was a better fit than the standard inflationary prediction somewhere in the parameter space. 2) If it turns out that the data are consistent with the standard inflationary prediction, it is important to know how large a difference from $n_\mathrm{t} = -r/8$ one would need before the data were able to discern it from the standard inflationary prediction.  

We assume the cosmological concordance model, i.e., the adiabatic $\Lambda$CDM model without tensor perturbations. On this playground, while varying all the standard $\Lambda$CDM parameters, we test a more complicated model that also has tensor perturbations.  If we assumed the inflationary consistency condition, $n_\mathrm{t} = -r/8$, we would have one extra parameter, $r$. A single value $r=0$ (which automatically also fixes $n_\mathrm{t}$ to zero) would mean that we recover the underlying simpler, tensorless, model. Any positive value means that the model is the more complicated one that has tensor perturbations.

However, once we relax the consistency condition, we cannot choose $r$ and $n_\mathrm{t}$ as our sampling parameters when finding parameter constraints against the data, if the data are consistent with $r=0$ or if this model is a relatively good fit to the data. Here the problem is that setting $r=0$ already reduces our model to the simpler, tensorless, model, no matter what value $n_\mathrm{t}$ has. Once we marginalize over the $n_\mathrm{t}$ direction of parameter space, e.g., in our Markov Chain Monte Carlo (MCMC) analysis, we get larger and larger weight the closer to zero $r$ is, since $n_\mathrm{t}$ can have here very large negative or positive values. In essence, how sharp a peak near zero $r$ in the posterior probability density we find, depends mainly on two things: how wide a prior range we allow for $n_\mathrm{t}$ and what bin size we use for plotting 1d pdf of $r$. With a fine binning, doubling the allowed range of $n_\mathrm{t}$ would roughly double the weight of the first bins in $r$. \emph{All this would lead to artificially tight constraints on $r$} (in the case where the data are consistent with $r=0$). As a result, we also would report over-optimistic constraining power for future experiments. 
We illustrate this $n_\mathrm{t}$-prior-range effect in Fig.~\ref{fig:ntprioreffect} using the BK15 data and keeping other than tensor parameters fixed to the \Planck\ best-fit $\Lambda$CDM values. The wider prior range we allow for $n_\mathrm{t}$, the closer to zero the tensor-to-scalar ratio at $k=0.01\,$Mpc$^{-1}$ will be. 
In addition, the posterior probability for $n_\mathrm{t}$ obtained by considering it as a primary parameter would depend on 
the pivot scale chosen
\footnote{
In a similar manner, there is a dependence on the pivot-scale chosen in fitting the isocurvature spectral index, as demonstrated in the $n_\mathrm{iso}$ panel of figure 21 of Ref.~\cite{Kurki-Suonio:2004bou}.}.


Fortunately, these problems with the 
combination $(r,n_\mathrm{t})$ can be overcome by a well-defined combination 
$\left[{\mathcal P}_{\rm t}(k_1),\, {\mathcal P}_{\rm t}(k_2)\right]$
or $(r_1,\,r_2)$,
where the former are the amplitudes of tensor perturbations at two different wavenumbers $k_1$ and $k_2$ and the latter are the tensor-to-scalar ratios at these two wavenumbers. With either of these parametrizations, the simpler, tensorless, model reduces to one single point $(0,0)$, instead of being an infinitely long line as in the $(r,n_\mathrm{t})$ case. The very same problem was identified 20 years ago when studying scalar isocurvature perturbations with a free spectral index in \cite{Kurki-Suonio:2004bou} and the suggested two-scale solution was implemented for the first time in \cite{Keskitalo:2006qv},
there named as an amplitude parametrization. Since then the method has also been applied to tensor perturbations by the \Planck\ Collaboration in \cite{Planck:2015sxf,Planck:2018jri}.

We adopt $(r_1,r_2)$ to describe the power-law spectrum of tensor modes \footnote{%
Equation~(\ref{matInter}) specifies a straight line in the $(\ln k,\, \ln {\cal P}_\mathrm{t})$ plane. This line goes through points $(\ln k_1,\, \ln {\cal P}_{\mathrm{t}1})$ and $(\ln k_2,\, \ln {\cal P}_{\mathrm{t}2})$, where ${\cal P}_{\mathrm{t}1} = {\cal P}_{\mathrm{t}}(k_1) = r_1 {\cal P}_{\mathrm{\cal R}} (k_1)$ and ${\cal P}_{\mathrm{t}2} = {\cal P}_{\mathrm{t}}(k_2) = r_2 {\cal P}_{\mathrm{\cal R}} (k_2)$.
%
},
\begin{align}
\begin{split}
{\cal P}_{\mathrm{t}}(k)=&\exp
\Biggl\{
\frac%
{\ln k  -\ln k_1}
{\ln k_2-\ln k_1}
\ln\bigl[ r_2 {\cal P}_{\mathrm{\cal R}} (k_2)\bigr]\\
&\qquad
-
\frac%
{\ln k  -\ln k_2}
{\ln k_2-\ln k_1}
\ln\bigl[ r_1 {\cal P}_{\mathrm{\cal R}} (k_1) \bigr]
\Biggr\}.
\label{matInter}
\end{split}
\end{align}
In order to obtain reasonable constraints on these tensor-to-scalar ratios at two different scales, the wavenumbers $k_1$ and $k_2$ should be chosen in such a way that $k_1$ corresponds roughly to the largest observable scale and $k_2$ to the smallest observable scale. 
The exact best choice could depend on the fraction of the sky observed and its coverage in multipoles (and therefore on wavenumbers). For this reason, we study in this paper different choices of the pair $(k_1,k_2)$, in particular, for future experiments.

In case of the tensor perturbations, there are two clearly observable features in the $B$-mode polarization at different $k$s: the reionization and recombination peaks. Thus we pick $k_1$ and $k_2$ near these features, respectively. Once we have used $(r_1,r_2)$ as the primary sampling parameters in our MCMC runs, we can calculate so-called derived parameters that have non-uniform priors (unlike the primary parameters). We often report $r$ at some middle scale between $k_1$ and $k_2$. If we want to show how the results would look like if the derived parameter had a uniform prior, we can simply weight our MCMC results by the inverse of the determinant of the Jacobian transform from the primary parameters to the derived one(s).
In particular, we compare many of our results by reporting $r_{0.01}$, i.e., $r$ at $k=0.01\,$Mpc$^{-1}$, which is close to the decorrelation scale of ($r$, $n_\mathrm{t}$).

However, any fundamental conclusions, such as a detection or determining whether $n_\mathrm{t} = -r/8$ is consistent with the data,should be drawn from the joint two-dimensional posterior distribution of $(r_1,r_2)$.
It should be kept in mind that even if $r_1$ and $r_2$ are drawn independently from a uniform distribution, i.e., $r_1$ and $r_2$ have flat priors in the MCMC runs, the individual posterior probability densities for $r_1$ alone, or for $r_2$ alone, or, in particular, for the derived parameter $r_{0.01}$ alone do not encode the full result. Instead, one should resort to the (marginalized) two-dimensional posterior of $(r_1,r_2)$ --- either its numerical or graphical representation.
For example, if the best-fit point and the 95\% CL contour in the $(r_1,r_2)$ plane were clearly away from point $(0,0)$, but the 99.7\% CL contour just reached $(0,0)$, then we might claim a weak $3\sigma$ detection.

\section{\Planck\ and BICEP/\Keck\ Array 2018 constraints\label{sec:realdata}}

\begin{figure*}
{
\includegraphics[width=0.50\textwidth]{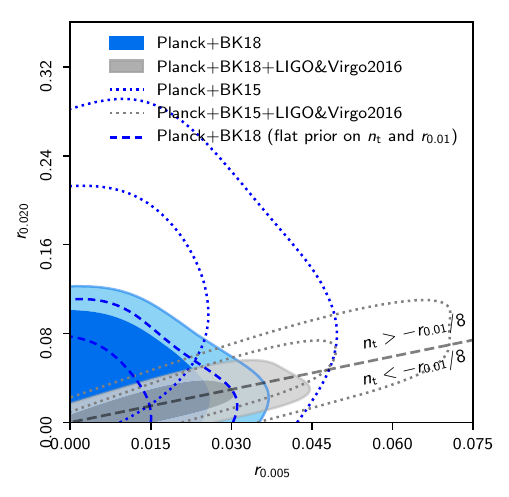}
\includegraphics[width=0.49\textwidth]{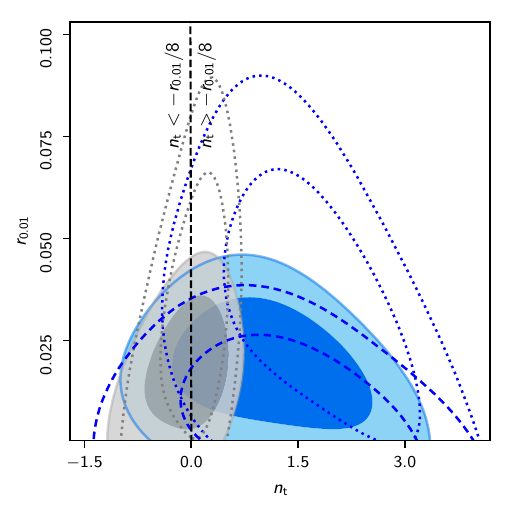}
\includegraphics[width=0.99\textwidth]{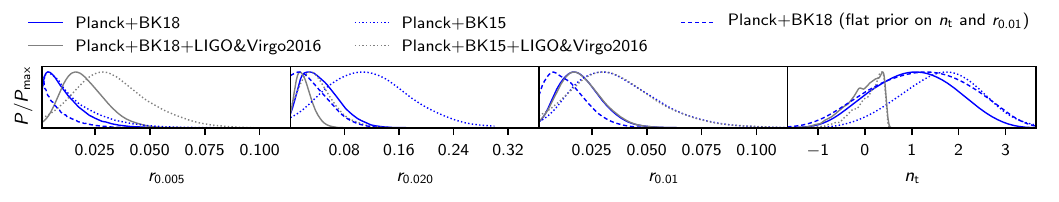}}
\caption{Posteriors with currently available real data for the tensor parameters when $n_\mathrm{t}$ is allowed to vary. The primary MCMC sampling parameters were the standard $\Lambda$CDM ones, the nuisance parameters of the likelihoods, and the (independent) tensor-to-scalar ratios at $k_1=0.005\,$Mpc$^{-1}$ and $k_2=0.02\,$Mpc$^{-1}$. The first panel shows 68\% CL and 95\% CL contours for the primary tensor parameters that have flat priors, except in the case of the blue-dashed curves that are obtained by mapping the MCMC run to indicate the results if flat priors on $n_\mathrm{t}$ and $r_{0.01}$ were employed instead. The second panel shows derived parameters $n_\mathrm{t}$ and $r_{0.01}$ that have non-flat priors, except again in the blue-dashed case. The last panel shows 1d marginalized results.
\label{fig:realdata}}
\end{figure*}

We now derive conservative constraints on primordial gravitational waves with the current data by adopting the two-scale parametrization described in the previous section. We use the \Planck\ 2018 data \citep{Planck:2019nip} and the latest BICEP/\Keck\ Array data release BK18 \citep{BICEP:2021xfz}.

We employ the \Planck\ 2018 baseline likelihood consisting of: a Gibbs sampling likelihood based on the component separated CMB map for temperature at $\ell \le 30$; $E$-mode simulation likelihood at $\ell \le 30$
based on the $100\times143\,$GHz cross angular power spectrum; Plik TTTEEE binned likelihood at high multipoles, i.e., $\ell > 30$.
We also include the lensing likelihood based on the four-point correlation function of the lensing signal in the conservative multipole range $8$--$400$.
As the BICEP/\Keck\ likelihood for $B$-mode polarization, we use the recently released likelihood which includes BICEP, \Keck\ Array and BICEP3 data up to the 2018 observing season \citep{BICEP:2021xfz}. We use \texttt{cosmomc} \cite{Lewis:2002ah,Lewis:2013hha} as the MCMC sampler and as a Bolzmann solver a modified version of \texttt{camb} \cite{Lewis:1999bs,Howlett:2012mh}, which includes the two-scale treatment for the tensor modes.

In this work, 
with the real data,
we use $k_1 = 0.005$ Mpc$^{-1}$ and $k_2 = 0.02$ Mpc$^{-1}$ and we also project
our results on $r$ at the scale $k=0.01$ Mpc$^{-1}$.
In \Planck\ X 2018 \cite{Planck:2018jri}, the use of $k_1 = 0.002$ Mpc$^{-1}$ was motivated by considering
one of the two most-used scales for the tensor-to-scalar ratio as a primary parameter, but here we instead prefer to use a slightly smaller scale $k_1 = 0.005$ Mpc$^{-1}$,
which has a broader overlap with the lowest multipoles probed by the BICEP/\Keck\ Array likelihood.

The 68\% CL and 95\% CL posterior constraints on our primary tensor parameters are shown by the blue shaded regions in the first panel of Fig.~\ref{fig:realdata}. For a comparison, we also show by blue dotted lines the constraints we obtained with an older BICEP/\Keck\ Array likelihood from 2015 (BK15) together with the \Planck\ 2018 data. BK18 data are consistent with no primordial gravitational waves also when relaxing $n_\mathrm{t}=-r/8$ or $n_\mathrm{t}=0$ and improve the constraints significantly over BK15 in combination with \Planck. As can be seen from the first panel of Fig.~\ref{fig:realdata}, the line $n_\mathrm{t}=-r/8$ is within the 68\% CL.

Using the same methodology as in \Planck\ X 2018 \cite{Planck:2018jri}, we also repeat the analysis by adding the LIGO\&Virgo 2016 95\% CL upper bound on the energy density parameter from gravitational waves, $\Omega_\mathrm{GW} < 1.7\times 10^{-7}$ at $k=(1.3$--$5.5)\times 10^{16}\,$Mpc$^{-1}$ \cite{LIGOScientific:2016jlg}, which is 18 orders larger $k$ than probed by the CMB $B$ mode. If the tensor power spectrum followed the strict power law that we assume, then a large region of positive $n_\mathrm{t}$ values would lead to a direct detection of stochastic primordial gravitational wave background that LIGO\&Virgo has not seen. The results, when making this huge extrapolation, are indicated by light gray in Fig.~\ref{fig:realdata}.

Finally, we reweigh our \Planck+BK18 MCMC chains to demonstrate that using $(r_{0.01},\,n_\mathrm{t})$ as primary parameters would artificially exaggerate the constraining power of the data by giving a large weight to the models that have $r$ near to zero (where $|n_\mathrm{t}|$ can be almost arbitrarily large and hence give extra weight to $r\sim 0$ upon marginalization). This case is indicated by the blue \emph{dashed} lines in Fig.~\ref{fig:realdata}.

The second panel of Fig.~\ref{fig:realdata} is based on the same analysis as the first panel, but now we show the derived parameters $r_{0.01}$ and $n_\mathrm{t}$ as in \cite{Planck:2015sxf,Planck:2018jri}, while the third panel shows the one-dimensional posterior probability densities (1d pdf) with peak values normalized to a same constant.

Our main result with the real data is
\begin{equation}
\left.\begin{aligned}
   r_{0.005} & < 0.030\\
   r_{0.02}  & <0.098 
\end{aligned}\ \right\}\ \
\mbox{\text{\parbox{4.9cm}{\begin{flushleft}
   (95\% CL, \Planck\ TT,TE,EE\\+lowE+lensing+BK18).
\end{flushleft}}}}
\label{eq:LCDMr1r2BK18}
\end{equation}
These constraints improve on the corresponding ones obtained with BK15, i.e. $r_{0.005} < 0.041$ and $0.009 < r_{0.02} < 0.23$. 
The constraints on the derived tensor parameters are
\mbox{$r_{0.01}< 0.039$} 
and $-0.61< n_{\mathrm t} < 2.73$ at 95\% CL, when using flat priors on the primary parameters. From the last two panels of Fig.~\ref{fig:realdata}, we notice that \Planck+BK18 gives by a factor of two a tighter constraint on $r_{0.01}$ compared to \Planck+BK15. 
Thus, BK18 represents a significant improvement also 
beyond the case of a fixed $n_\mathrm{t}$ studied in \cite{BICEP:2021xfz}. Naturally, 
the 95\% CL contours on $n_\mathrm{t}$ do not improve since BK18 brings the constraint on the actual tensor contribution closer to zero.

The mean of the posterior at $n_\mathrm{t} \sim 1$ is due to the transfer function of primordial gravitational waves (that strongly damps their contribution to $C_\ell^{BB}$ at $\ell \gtrsim 200$ unless the primordial tilt $n_\mathrm{t}$ is very large), in combination with the CMB lensing, noise plus foregrounds, and cosmic variance. 
The primordial signal which minimizes 
the $\chi^2$ have (the amplitude and) a tensor tilt that mimics the effective noise.
This phenomenon is analogous to the apparent preference of $n_\mathrm{iso} \sim 3$ for the CDM isocurvature perturbations in the lack of a detection of such a component, as explained, e.g., in Refs.~\cite{Kurki-Suonio:2004bou,Valiviita:2009bp,Valiviita:2012ub}. 

\begin{figure}
\includegraphics[width=0.48\textwidth]{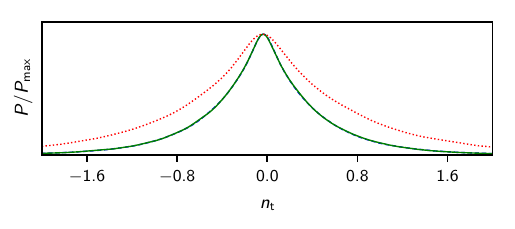}
\includegraphics[width=0.48\textwidth]{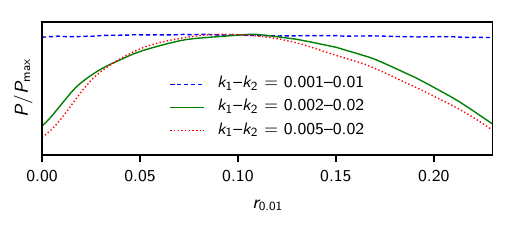}
\caption{The induced priors on the derived tensor parameters $n_\mathrm{t}$ and $r_{0.01}$ when sampling $r_1$ and $r_2$ from a flat prior [0,0.25]. Three different choices of $(k_1,k_2)$ are shown. 
The width of the induced prior for $n_\mathrm{t}$ depends on $k_2/k_1$, being the wider the closer $k_1$ is to $k_2$: for the first two choices the prior width is $\sigma(n_\mathrm{t}) \approx 0.6$, and for the last choice, $\sigma(n_\mathrm{t}) \approx 1$. Note that these induced priors do not use any likelihood/data.
\label{fig:ntprior}}
\end{figure}

The flat priors on $r_1$ and $r_2$ induce a non-flat prior on $n_\mathrm{t}$ \cite{Galloni:2022mok} with a peak at $n_\mathrm{t} \approx 0$, as shown in the upper panel of Fig.~\ref{fig:ntprior}. This might introduce a mild push on $n_\mathrm{t}$ toward zero, but in Sec.~\ref{sec:nullcase} we show by using simulated $r=0$ data that this push does not outweigh the above-mentioned natural preference of $n_\mathrm{t} \sim 1$ in the null case when using the CMB data alone. 
The symmetric posterior of $n_\mathrm{t}$ around $1$ is an implication of \Planck+BK18 being consistent with no tensors within the sensitivity of these data.
The induced prior on $r_{0.01}$ (see the red dotted curve in the lower panel of Fig.~\ref{fig:ntprior}) mildly pushes $r_{0.01}$ away from zero, making our quoted upper bound a conservative one
\footnote{As the two-dimensional analysis of $(r_1,\,r_2)$ does not indicate any detection of a non-zero tensor contribution, i.e., the best fit is very near to $(0,\,0)$ and $(0,\,0)$ is in the 68\% CL region, we report the conservative  95\% CL upper bounds on the tensor-to-scalar ratio by forcing a one-tail analysis in \texttt{getdist}.}.

In Table~\ref{tab:cosmoparamsrealdata}, 
we do not find any statistically significant shift in the remaining cosmological parameters when the consistency relation between the tensor-to-scalar ratio and the tensor tilt is relaxed. We also do not observe major degeneracy among $r_{0.005}$, $r_{0.02}$, and the foreground/nuisance parameters of the BK18 likelihood in combination with \Planck. 
When $n_\mathrm{t}$ is allowed to vary, the low-$k$ constraint, $r_{0.005} < 0.03$, does not degrade compared to the derived constraint of the $n_\mathrm{t} = -r/8$ case. Indeed, as there is more allowed parameter-space volume at the positive $n_\mathrm{t}$, the low-$k$ constraint is slightly tighter than in the $n_\mathrm{t} = -r/8$ model. Once we pass the recombination bump, the data become less and less sensitive to the primordial tensor modes as they are damped by the transfer function. This is reflected by the fact that the constraint on $r_{0.02}$ is by a factor of three weaker than the corresponding bound when keeping $n_\mathrm{t}$ fixed. Finally, once projected on the standard pivot scale, $k_* = 0.05$ Mpc$^{-1}$, we have $r_{0.05} < 0.71$ at 95\% CL when $n_\mathrm{t}$ is allowed to vary, which is by nearly a factor of 20 weaker than the upper bound $0.035$ obtained  with a fixed $n_\mathrm{t}=-r/8$.
\begin{table}
\begin{ruledtabular}
\begin{tabular}{lccdr}
\textrm{Parameters}&
\textrm{\Planck+BK18}&\textrm{\Planck+BK18} \\
&$n_{\rm t}=-r/8$ & free $n_{\rm t}$\\ 
\colrule
\noalign{\vskip 2pt}
$\Omega_\mathrm{b} h^2$ & $0.0224\pm 0.0001$ & $0.0224 \pm 0.0001$  \\ $\Omega_\mathrm{c} h^2$ & $0.120\pm 0.001 $ &  $0.120\pm 0.001$ \\ 
100 $\theta$ & $1.
0409 \pm 0.0003$ & $1.0409 \pm 0.0003$\\ 
 $\tau$ & $0.0546_{-0.0072}^{+0.0073}$& $0.0544\pm 0.0073$ \\ 
 $\ln(10^{10} A_\mathrm{s})$ & $3.045\pm 0.014$ & $3.044 \pm 0.014$ \\ 
 $n_\mathrm{s}$ & $0.9653\pm 0.0041 $ & $0.9656\pm 0.0041$ \\ 
 \noalign{\vskip 2pt}
 \colrule
 \noalign{\vskip 2pt}
 $r_{0.005}$ & \textcolor{gray}{$(<0.032)$} & $\mathbf{<0.030}$ \\
 $r_{0.02}$ & \textcolor{gray}{$( <0.034)$} & $\mathbf{<0.098}$ \\
 $r_{0.05}$ & $\mathbf{<0.035}$ & \textcolor{gray}{$(<0.71)$} \\
\end{tabular}
\end{ruledtabular}
\caption{\label{tab:cosmoparamsrealdata} 68\% CL constraints for the $\Lambda$CDM parameters
with \Planck+BK18 data.
In bold are the 95\% CL upper bounds for the primary tensor parameters, and
in parenthesis are the derived tensor parameters.
} 
\end{table}

\section{Forecasts for future experiments\label{sec:forecasts}}

In the next years, there will be several experiments 
devoted to CMB polarization measurements and, in particular, to the $B$ modes \citep{SimonsObservatory:2018koc,LSPE:2020uos,LiteBIRD:2022cnt,CMB-S4:2016ple}.
In this section, we compute forecasts 
when $n_\mathrm{t}$ is allowed to vary  
by using the two-scale parametrization (see also \cite{CORE:2016ymi}) and simulated $B$-mode data representative of the future CMB measurements, taking as an example the Lite
(Light) satellite for the study of $B$-mode polarization and Inflation from cosmic background
Radiation Detection (LiteBIRD) \citep{LiteBIRD:2022cnt}, selected by the Japan Aerospace Exploration Agency (JAXA) as a strategic large class mission to which, in addition to Japan, also Europe, the United States, and Canada contribute.

\begin{table}
\begin{tabular}{ p{1.95cm} p{1.95cm} p{1.95cm} p{1.95cm}   }
 \hline
 \hline
 \noalign{\vskip 3pt}
 \multicolumn{4}{c}{LiteBIRD} \\
 \hline
 \noalign{\vskip 3pt}
Frequency & T-sens & P-sens &FWHM \\
(GHz) &  ($\mu K\,$arcmin) & ($\mu K\,$arcmin) & (arcmin)\\
\noalign{\vskip 2pt}
\noalign{\hrule\vskip 3pt}
78 & 8.53 &12.07&36.9\\
89 &7.99 &11.30& 33.0\\
100 &4.64 &6.56& 30.2\\
119 &3.24 &4.58& 26.3\\
140 &3.39 &4.79& 23.7\\
166 &3.94 &5.57& 28.9\\
195 &4.14 &5.85& 28.0\\
\hline
\hline
\end{tabular}
\caption{\label{tab:LB}A LiteBIRD-like configuration  of the instrument central frequency channels,
following the characterization given in \cite{LiteBIRD:2022cnt}. } 
\end{table}
\begin{figure}
\includegraphics[width=0.48\textwidth]{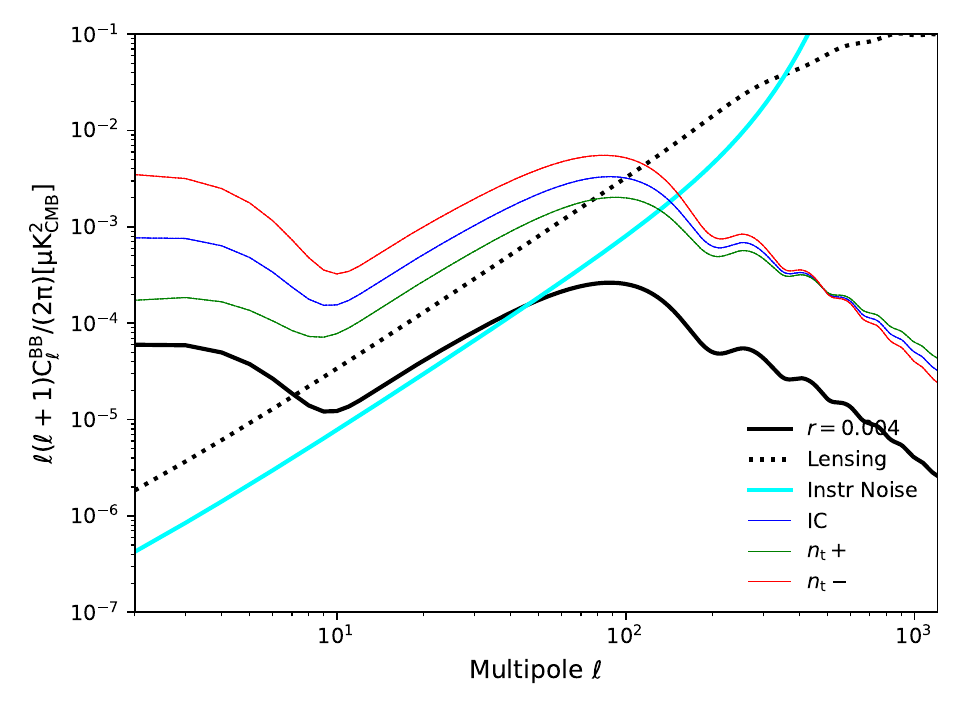}\\
\caption{\label{fig:APS}
The CMB $B$-mode anisotropy angular power spectra resulting from the primordial tensor perturbations for three fiducial models with $r_{0.05}=0.05$ are represented in three different colors (blue, green red). 
The dotted black curve is the CMB lensing signal
and the cyan curve is the instrumental noise of a LiteBIRD-like configuration. 
}
\end{figure}
\begin{figure*}
    \centering
    \includegraphics[width=0.3\textwidth]{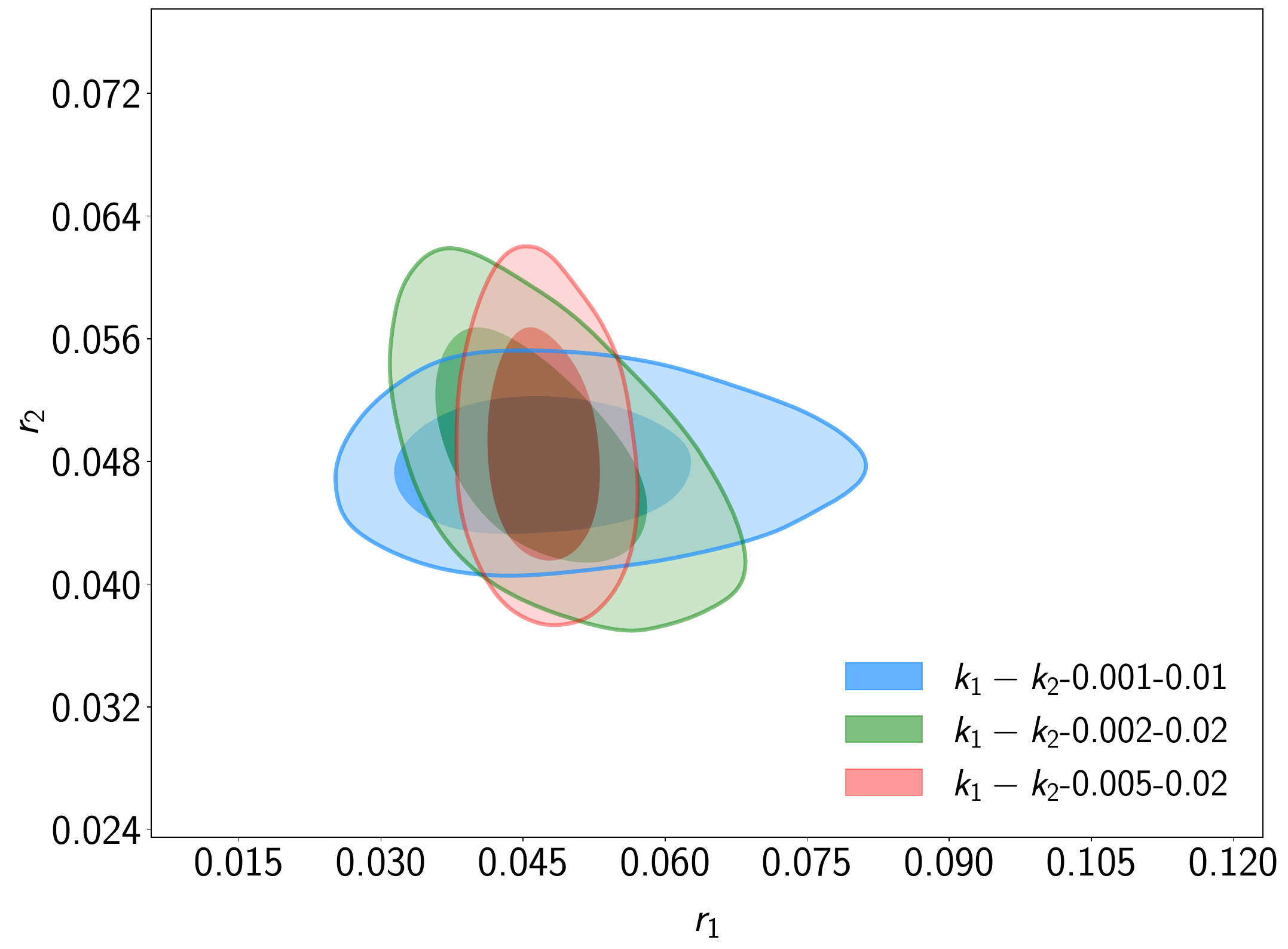}
    \includegraphics[width=0.3\textwidth]{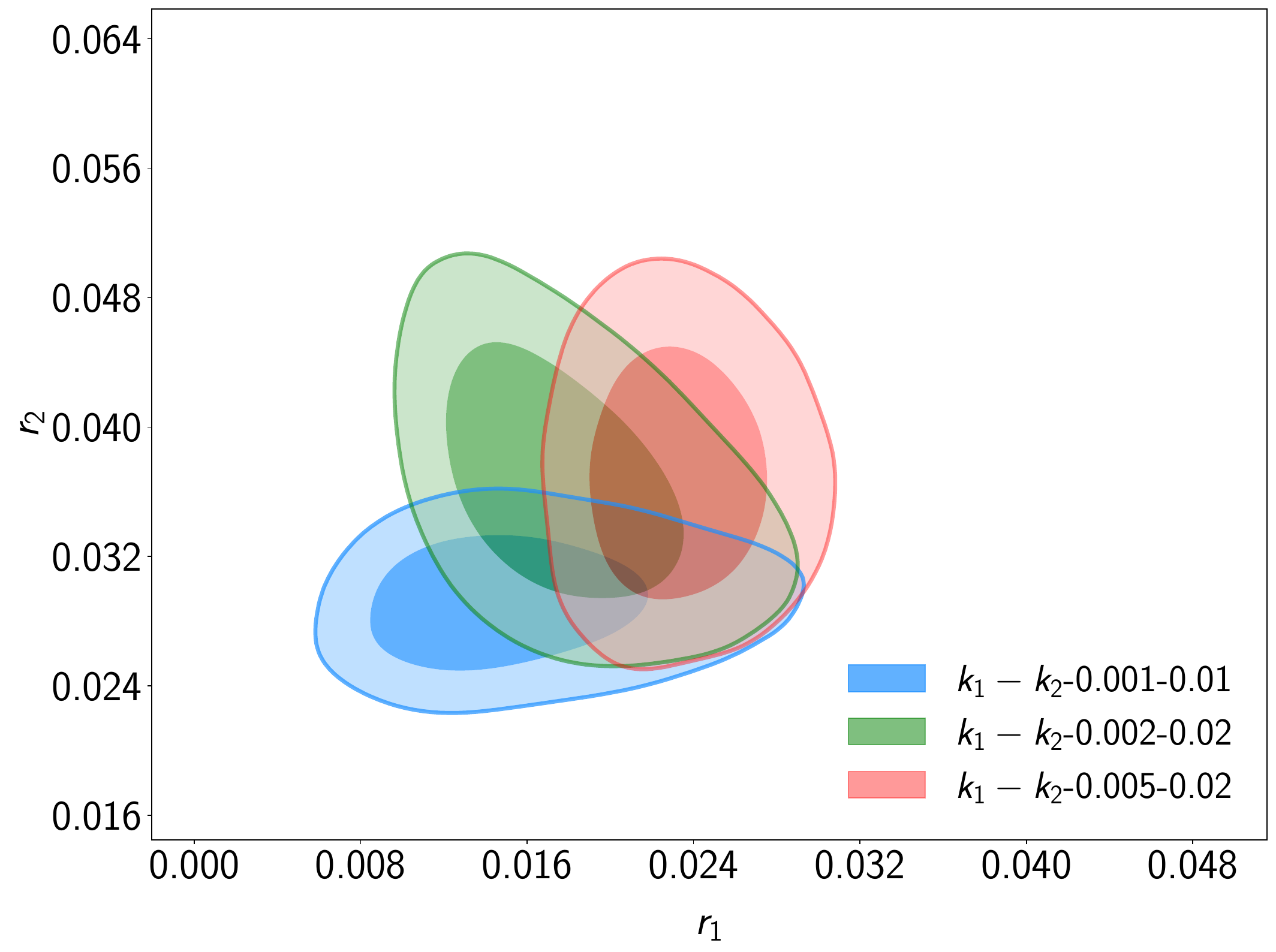}
    \includegraphics[width=0.3\textwidth]{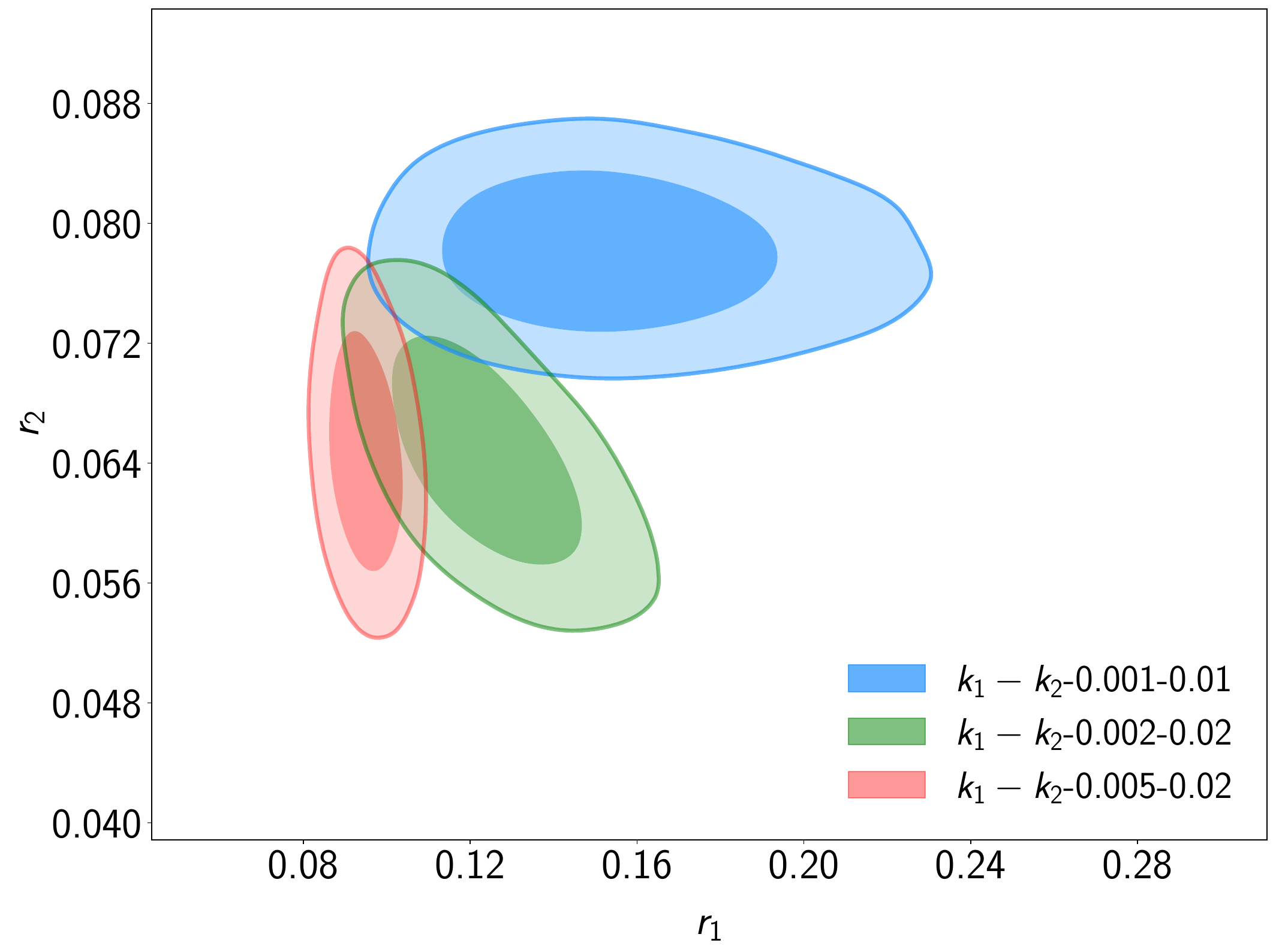}
    \caption{Two-dimensional posterior distributions for $r_1$ and $r_2$, assuming a common $r_{0.05}=0.05$ as a fiducial model for the simulated data. The panels are from left to right: the inflationary consistency (IC), the positive tensor tilt ($n_\mathrm{t}+$), and the negative tensor tilt  ($n_\mathrm{t}-$). Note that with different choices of $k_1$---$k_2$ the tensor-to-scalar ratios $r_1$ and $r_2$ represent $r$ at different scales, and hence, e.g., the areas covered by the 95\% CL posterior are not directly comparable in this figure.}
    \label{fig:largefid_r1_r2}
\end{figure*}
\begin{figure*}
\includegraphics[width=0.46\textwidth]{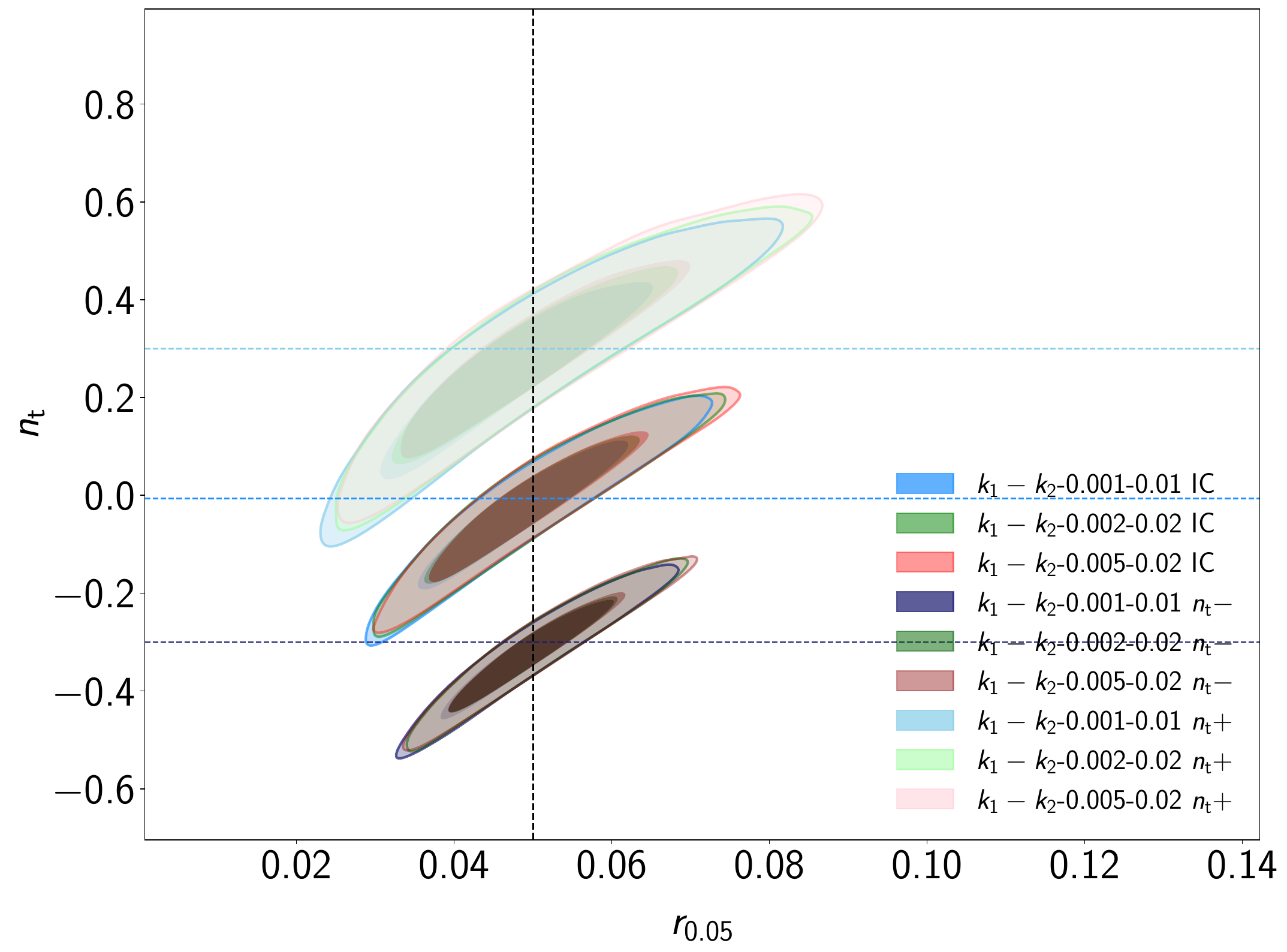}\includegraphics[width=0.46\textwidth]{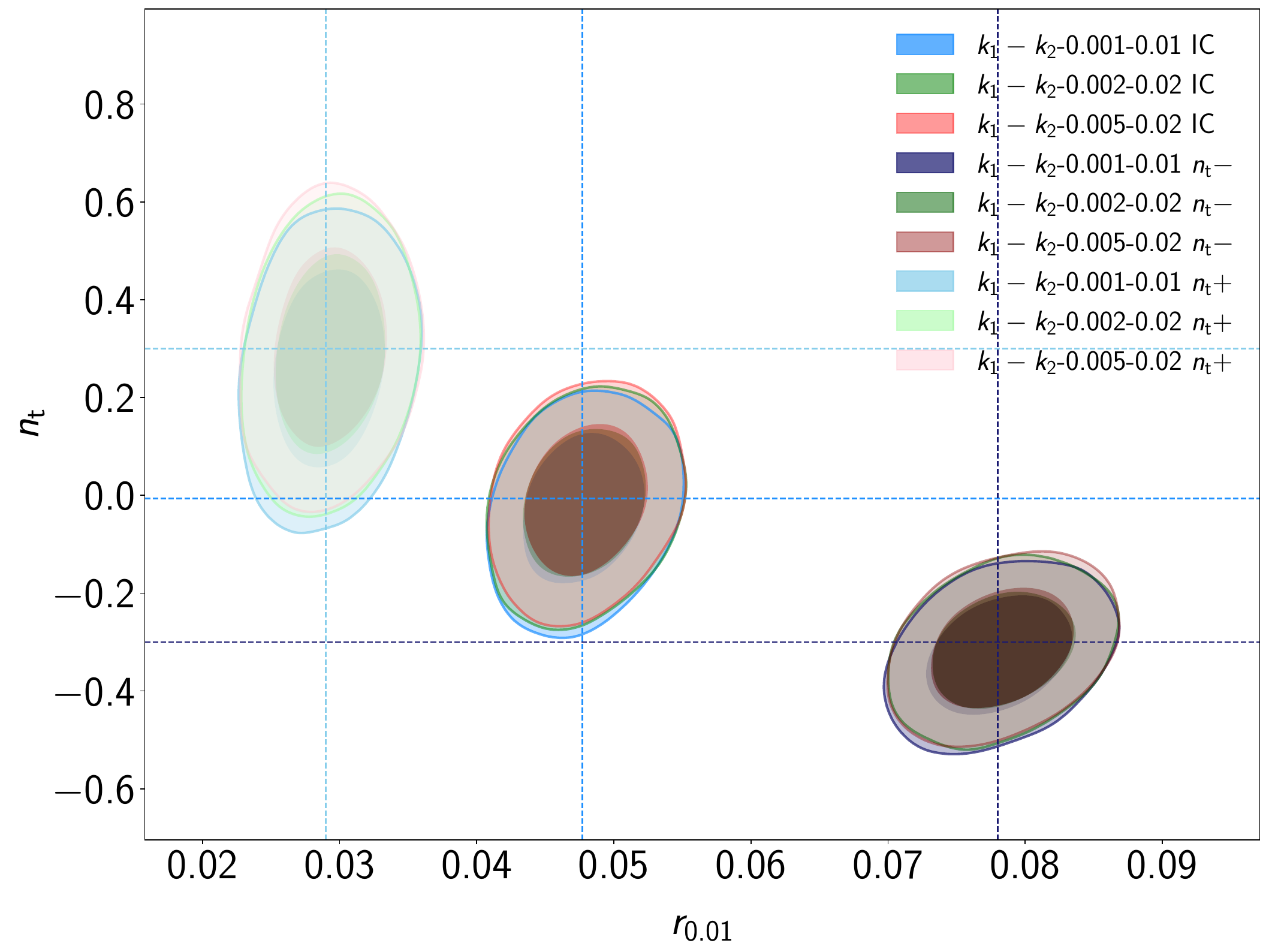}
\caption{Two-dimensional posterior distributions for the derived parameters $r_{0.05}$ and $n_{\rm t}$ and for $r_{0.01}$ and $n_{\rm t}$ for the three cases assuming a common $r_{0.05}=0.05$.}
\label{fig:2D05}
\end{figure*}

\subsection{Simulated data and methodology}

We consider LiteBIRD-like instrumental specifications given in Table~\ref{tab:LB}.
We produce simulated data for $T,E$ by considering the inverse noise weighting of the central frequency channels and by assuming that the lowest  and highest frequencies are used to separate the foreground emission as done in \cite{LiteBIRD:2022cnt} (see also \cite{CORE:2016ymi}). For the $B$-mode polarization (in addition to the instrumental noise), we include the following two sources of confusion: the lensing signal and a contribution which mimics the foreground residuals.
These inputs are inserted in a Wishart-like likelihood with an effective sky fraction of 70\% (60\%) for $T,E$ ($B$).
With these settings, we obtain for the $n_\mathrm{t}=-r/8$ model a $\sigma(r) \sim 0.0013$, which is 30\% larger than the LiteBIRD value \cite{LiteBIRD:2022cnt} which includes systematic effects. 
Our constraints can therefore be seen as a conservative assessment of the LiteBIRD capabilities.

We create simulated data from three fiducial models. Our first fiducial model satisfies the inflationary consistency condition $n_\mathrm{t} = -r/8$ (IC) and is motivated by SSSRI with Bunch-Davies quantum initial condition. 
As a second case, we consider a positive value for the tensor tilt ($n_\mathrm{t}+$), as occurs when the null-energy condition during inflation is violated \cite{Baldi:2005gk} or in Galileon inflation \cite{Kobayashi:2010cm}. We fix $n_\mathrm{t}=0.3$, which 
is allowed by the \Planck\ constraints on primordial non-Gaussianity \cite{2011PhRvD..83j3524K,Planck:2015zfm,Planck:2015sxf,Planck:2019kim} in Galileon inflation \cite{Kobayashi:2010cm}.
As a third case, we consider $n_{\rm t}=-r/(8 c_\mathrm{s})$, theoretically predicted by slow-roll inflation with a non-canonical  kinetic term, which leads to a non-trivial speed of sound $(0 < )\ c_\mathrm{s} < 1$, with $c_\mathrm{s} = \partial p (\phi, X)/\partial X$. 
Here $p$ is the Lagrangian for the inflaton and $X = - \partial_\mu \phi \partial^\mu \phi/2$. As a value for the inflaton speed of sound we consider the 95\% CL lower limit $c_\mathrm{s} = 0.02$ obtained by the constraint on primordial non-Gaussianity \cite{Planck:2015zfm,Planck:2019kim}, which is enhanced by the non-trivial speed of sound. This value leads to a negative value for the tensor tilt, i.e. $n_\mathrm{t} = -0.3$, and we denote this case by $n_\mathrm{t}-$.

We set $r=0.05$ (at $k=0.05\,$Mpc) in all three fiducial models and assume the underlying $\Lambda$CDM cosmology to be consistent with the \Planck\ 2018 baseline results: $\Omega_\mathrm{b} h^2$ = 0.02237,
$\Omega_\mathrm{c} h^2$ = 0.120, $100\,\theta$ = 1.04092,
$\tau$ = 0.0544, $n_\mathrm{s}$ = 0.9649, $\ln(10^{10}A_\mathrm{s})$ = 3.044. Figure~\ref{fig:APS} shows the $B$-mode angular power spectra for these fiducial models and, for comparison, the instrumental noise and lensing signal.

With each of the three simulated data sets, we run three separate full MCMC runs (i.e.,  nine runs in total), but choosing three different pairs $k_1$---$k_2$ for the two-scale parametrization. As explained in Sec.~\ref{sec:two-scale}, the best choice depends on the $k$-coverage (multipole coverage) of the data. In addition, the best choice may depend on the actual underlying model to be recovered. In this subsection our fiducial models described above are chosen to have a largish $r$ and/or $n_\mathrm{t}$ (allowed by the previous BK15 data release), in order to test/exaggerate the latter effect. In the next subsection we repeat the analysis using fiducials that would be allowed by the current constraints.

We test the sensitivity of parameter estimation to the choice of $k_1$ and $k_2$ by using the pairs $0.001$---$0.01$, $0.002$---$0.02$, and $0.005$---$0.02$ Mpc$^{-1}$. All these scales effectively correspond to multipoles where the expected tensor signal is non-negligible, differently from the conventional $k_*=0.05\,$Mpc$^{-1}$, where the signal is damped by the tensor transfer function. 

We check the posteriors of the primary tensor sampling parameters $r_1$ and $r_2$ in Fig.~\ref{fig:largefid_r1_r2}.
We note that for all three fiducial models the green case ($k_1$---$k_2$=$0.002$---$0.02$) performs worst since it leads to a degeneracy between $r_1$ and $r_2$ for this LiteBIRD-like configuration that we use, which degrades the determination of both these parameters. 

The derived tensor parameters from the same MCMC runs as above are shown in Fig.~\ref{fig:2D05}. 
We notice that for all three fiducial models considered, any of our choices of $k_1$---$k_2$ leads to an excellent recovery of the derived parameters $r_{0.05}$, $r_{0.01}$, and $n_\mathrm{t}$ in terms of the median of the posterior. From Fig.~\ref{fig:2D05}, we see that distinguishing $n_{\rm t}=-r/8$ from the exact scale invariance is out of reach as expected \cite{Knox:2002pe}. 
For the modified sound speed case, $n_\mathrm{t}-$ in the darkest colors, we observe significantly better constraints on the tensor spectral index. This result reflects the fact that we used a fixed $r_{0.05}=0.05$ as an input fiducial, which, in the $n_\mathrm{t}-$ case, translates into a large $r$ at the LiteBIRD sensitivity region, as is obvious from the values of $r_{0.01}$ in the right panel of Fig.~\ref{fig:2D05}. Having a large $r$ naturally leads to tighter constraints on $n_\mathrm{t}$. This should be kept in mind when interpreting the results. For the $n_{\rm t}=-r/8$ case, it does not make a big qualitative difference on what scale one quotes $r$, as $n_{\rm t} \approx 0$ (and $n_\mathrm{s}-1 \approx -0.04$). However, once we relax the consistency condition, we are required to be careful and explicit with the scales. What matters is $r$ in the sensitivity region of the experiment. At the standard pivot scale, $k=0.05\,$Mpc$^{-1}$, the tensor-to-scalar ratio can then be very large (small) in the $n_{\rm t}+$ ($n_{\rm t}-$) case.

\subsection{Forecasts for realistic cases}

We now present the forecasts by choosing fiducial models in such a way that $r_{0.01}$ and $n_\mathrm{t}$ are inside the 95\% CL region of \Planck+BK18 posterior.
All other aspects of the analysis stay the same as in the previous subsection. 

\subsubsection{Inflationary consistency (IC)}

\begin{figure*}
\includegraphics[width=0.32\textwidth]{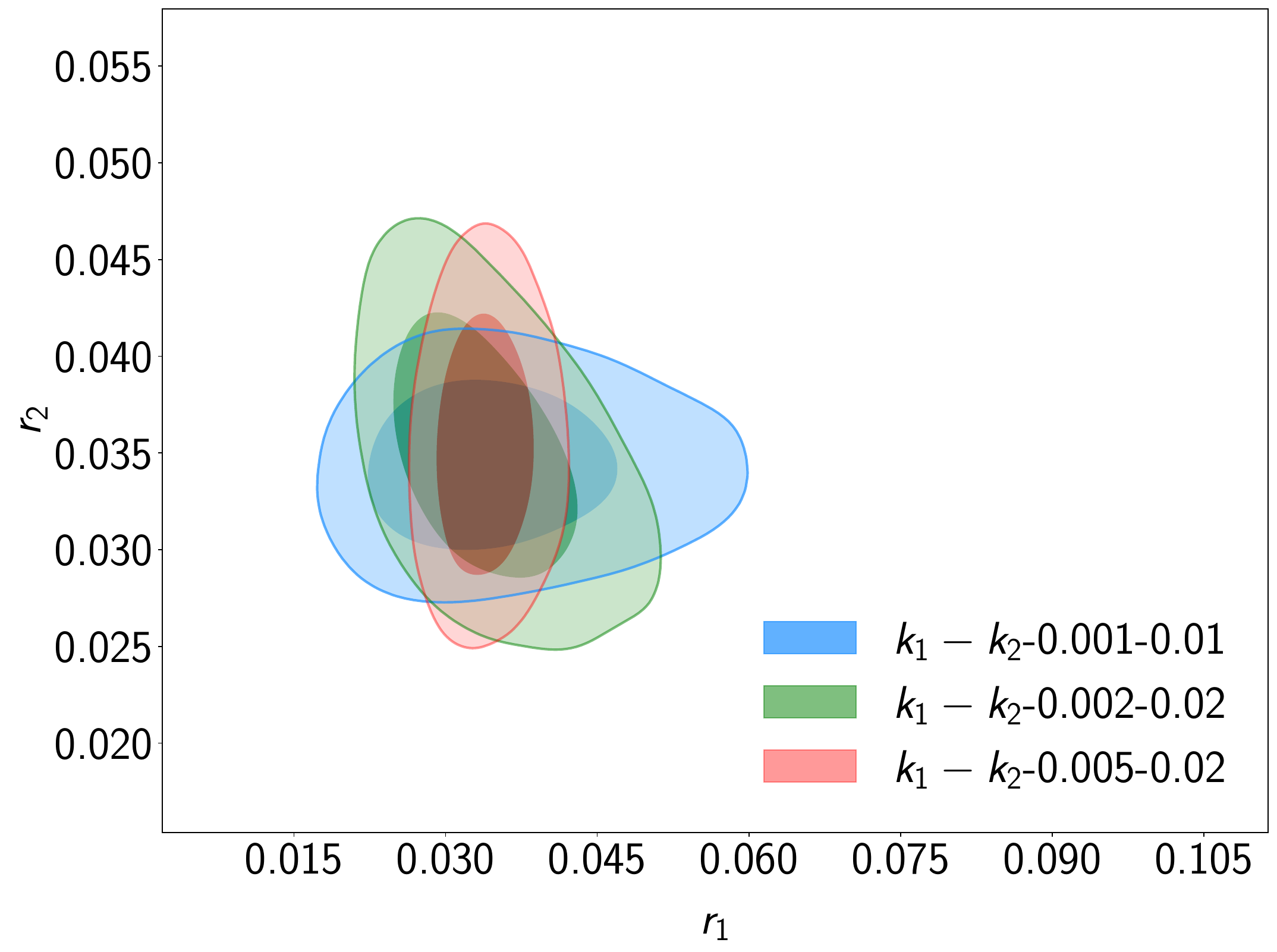}
\includegraphics[width=0.32\textwidth]{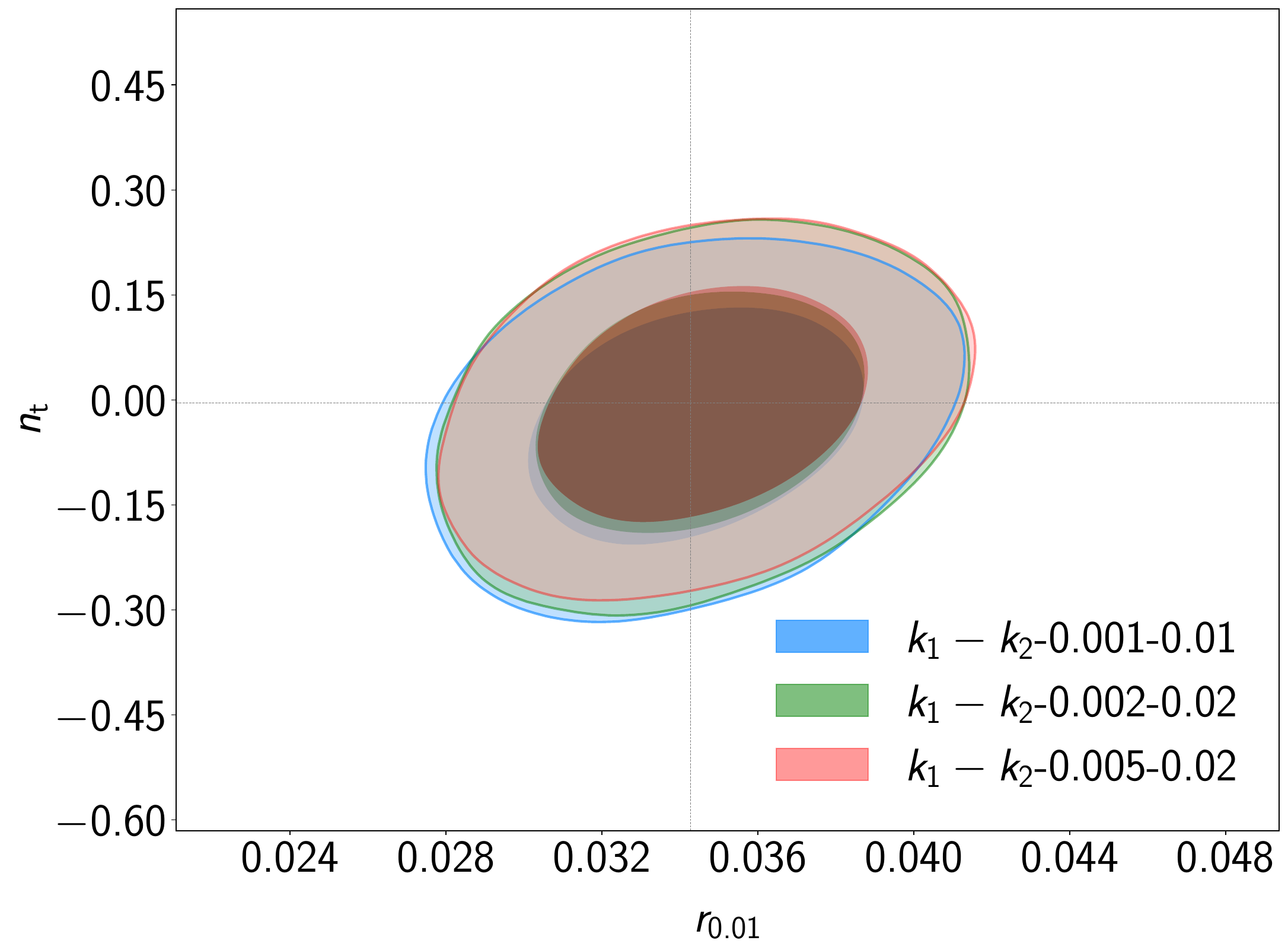}
\includegraphics[width=0.32\textwidth]{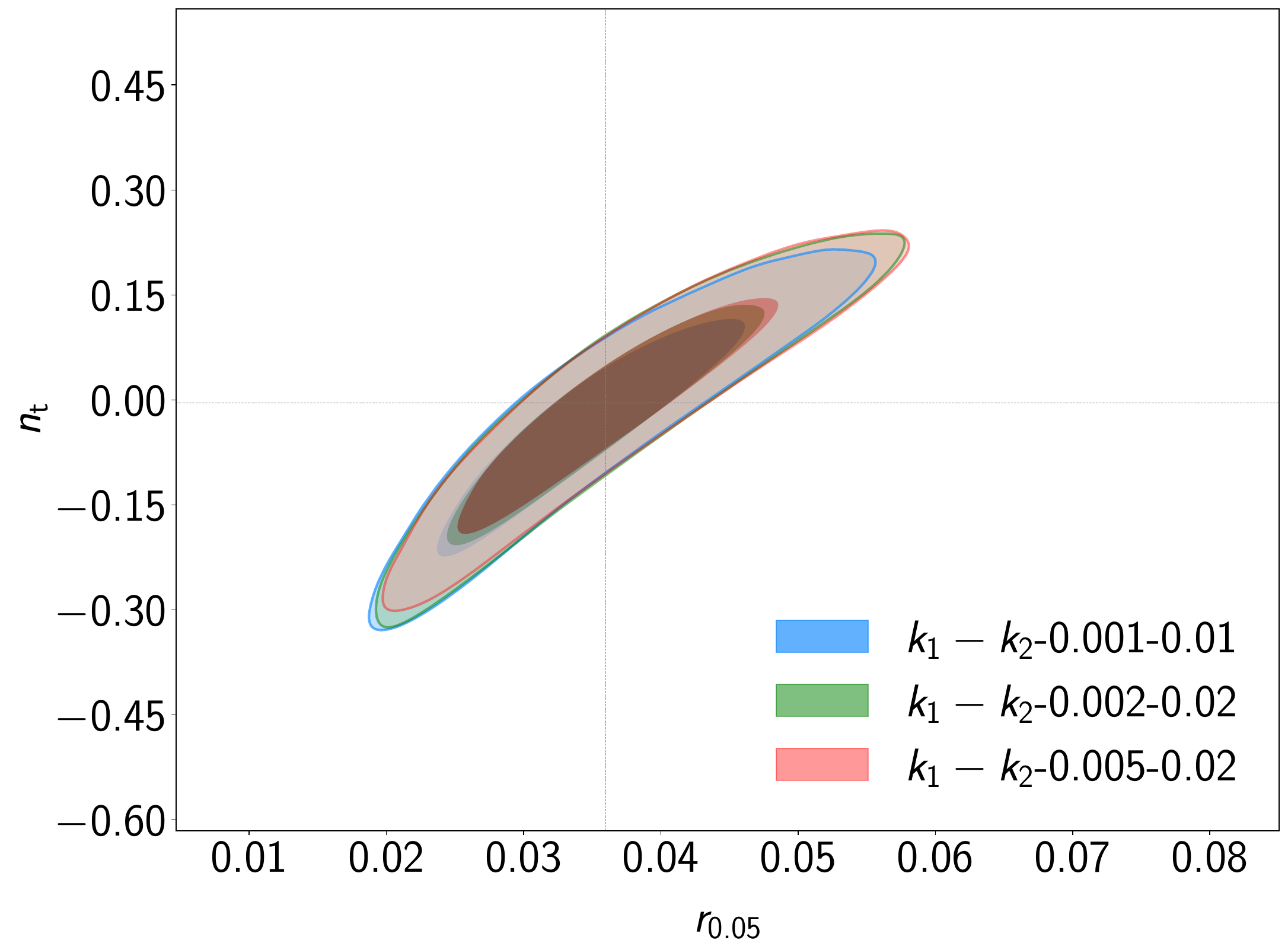}
\caption{Realistic IC case forecasts for the marginalized $(r_1,r_2)$ (left panel), $(r_{0.01},n_{\rm t})$ (middle panel), $(r_{0.05},n_{\rm t})$ (right panel) contours obtained by a two-scale analysis with three different pairs of scales (in blue, green, and red), using simulated LiteBIRD-like data with tensor parameters of the fiducial model having the values $r_{0.05}=0.036$ and
$n_{\rm t}=-0.036/8=-0.0045$. Note that the $r$ axes do not start from zero in any of the panels.
}
\label{fig:r0p036_1}
\end{figure*}

We start with the inflationary consistency case where we assume $r_{0.05}=0.036$ (giving $n_\mathrm{t} = -0.0045$), compatible with the 95\% CL region by the \Planck+BK18 data. The constraints for the tensor parameters are presented in Table \ref{tab:IC036} and the posteriors in Fig.~\ref{fig:r0p036_1}. 
Again the area covered by the two-dimensional contours for the primary parameters $r_1$ and $r_2$ apparently depends on the choice of scales, but this is due to plotting $r$s at different scales on the same figure. The dependence disappears when projecting on an amplitude ($r_{0.01}$ or $r_{0.05}$) and tilt in the middle and last panels. The first panel indicates that the green case ($k_1$---$k_2$=$0.002$---$0.02$) has a  degeneracy between $r_1$ and $r_2$. Table~\ref{tab:IC036} shows that marginally the best recovery of the input parameters is achieved by $k_1$---$k_2$=$0.001$---$0.01$. The measurement precision in this case is $\sigma(r_{0.001})\approx0.009$ and $\sigma(r_{0.01})\approx0.003$, which implies approximately a $10\,\sigma$ detection of our non-zero input $r_{0.01}=0.0343$. 

\renewcommand{\dblfloatpagefraction}{0.3}
\begin{figure*}
\includegraphics[width=0.33\textwidth]{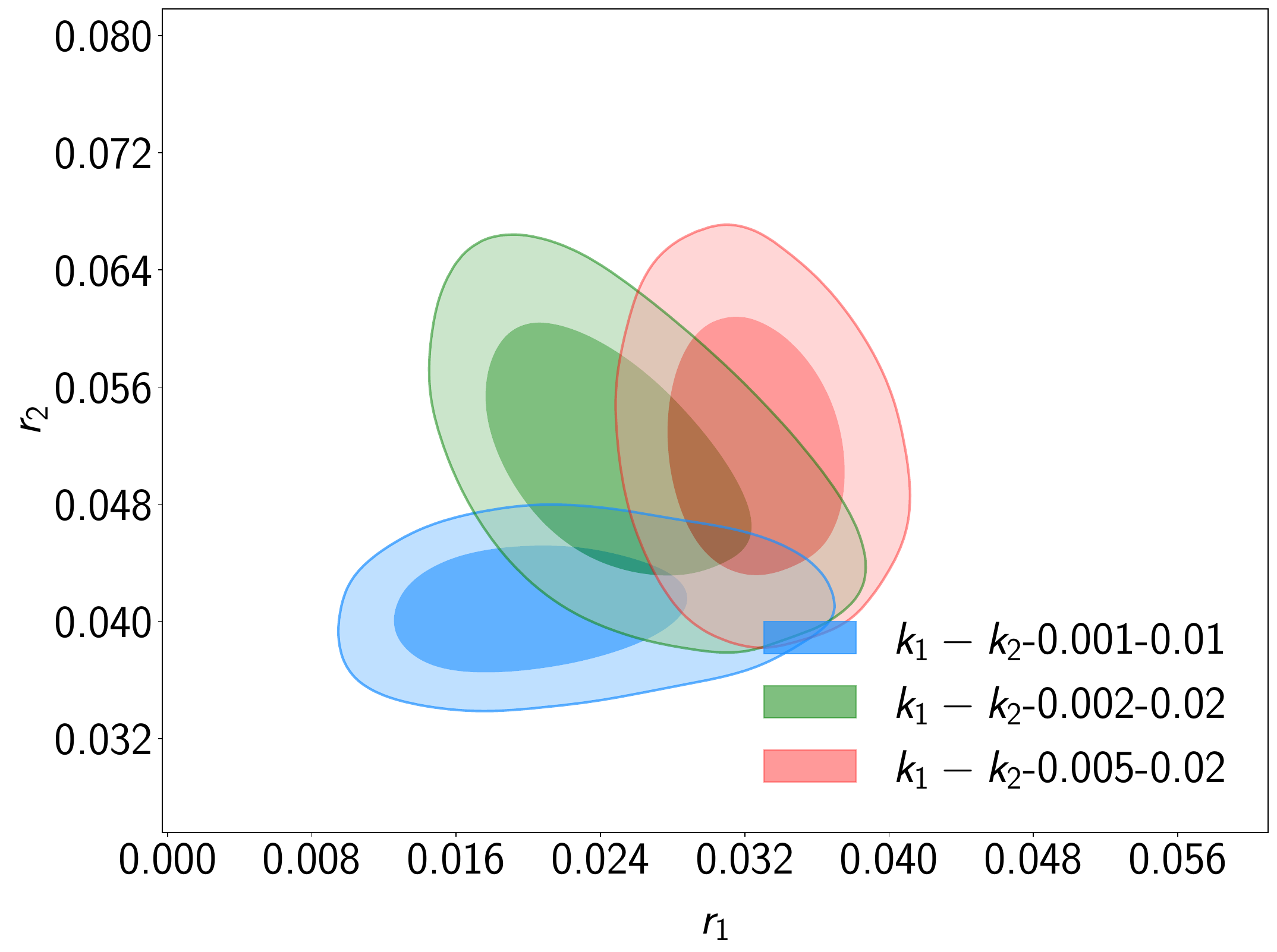}\includegraphics[width=0.33\textwidth]{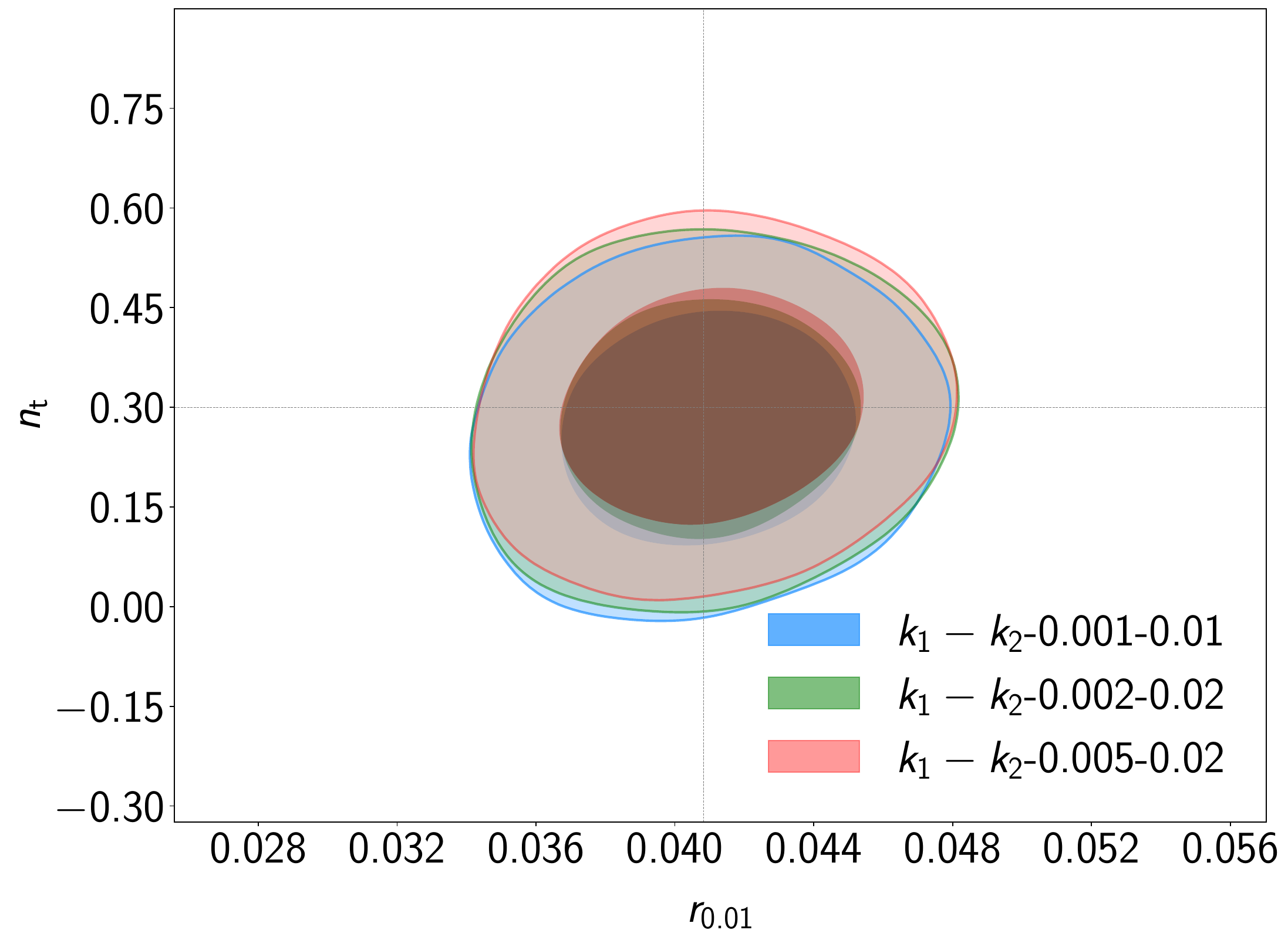}
\includegraphics[width=0.33\textwidth]{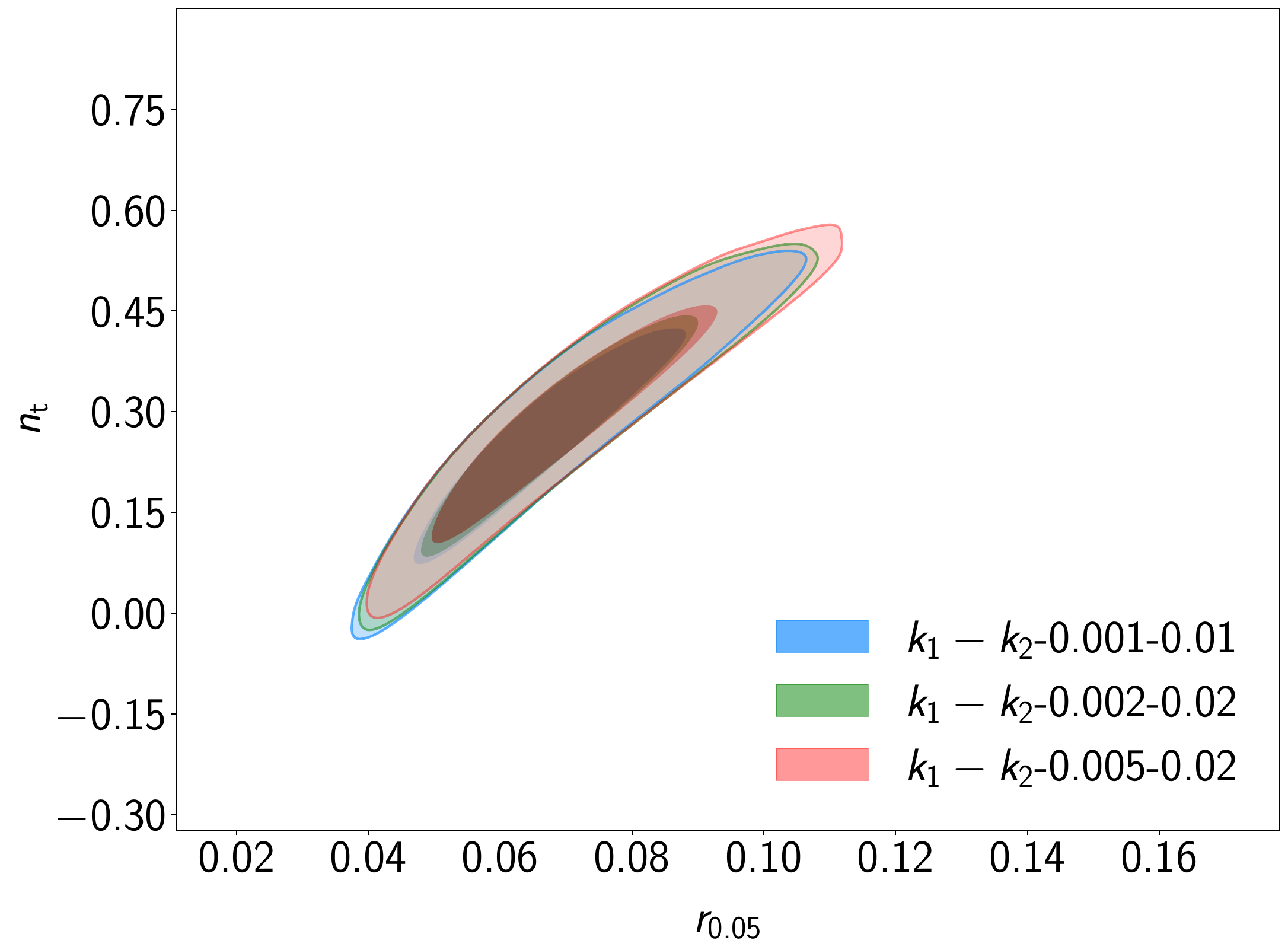}
\caption{Realistic $n_\mathrm{t}+$ case forecasts for the marginalized $(r_1,r_2)$ (left panel), $(r_{0.01},n_{\rm t})$ (middle panel), $(r_{0.05},n_{\rm t})$ (right panel) contours obtained by a two-scale analysis with three different pairs of scales (in blue, green, and red), using simulated data based on the fiducial tensor parameters $r_{0.05}=0.07$ and
$n_{\rm t}=0.3$. 
}
\label{fig:fig:r0p07_1}
\end{figure*}
\begin{figure*}
\includegraphics[width=0.33\textwidth]{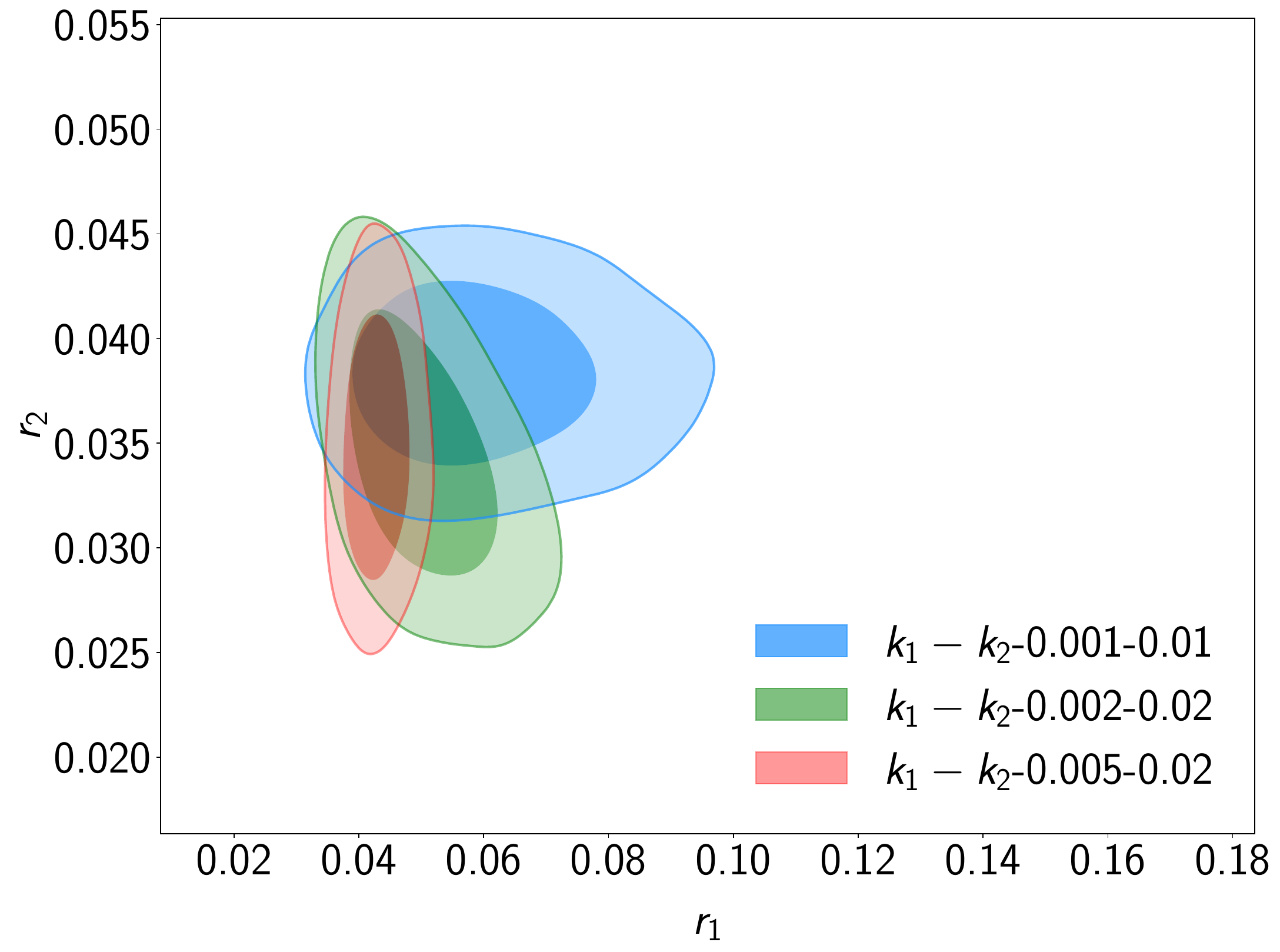}\includegraphics[width=0.33\textwidth]{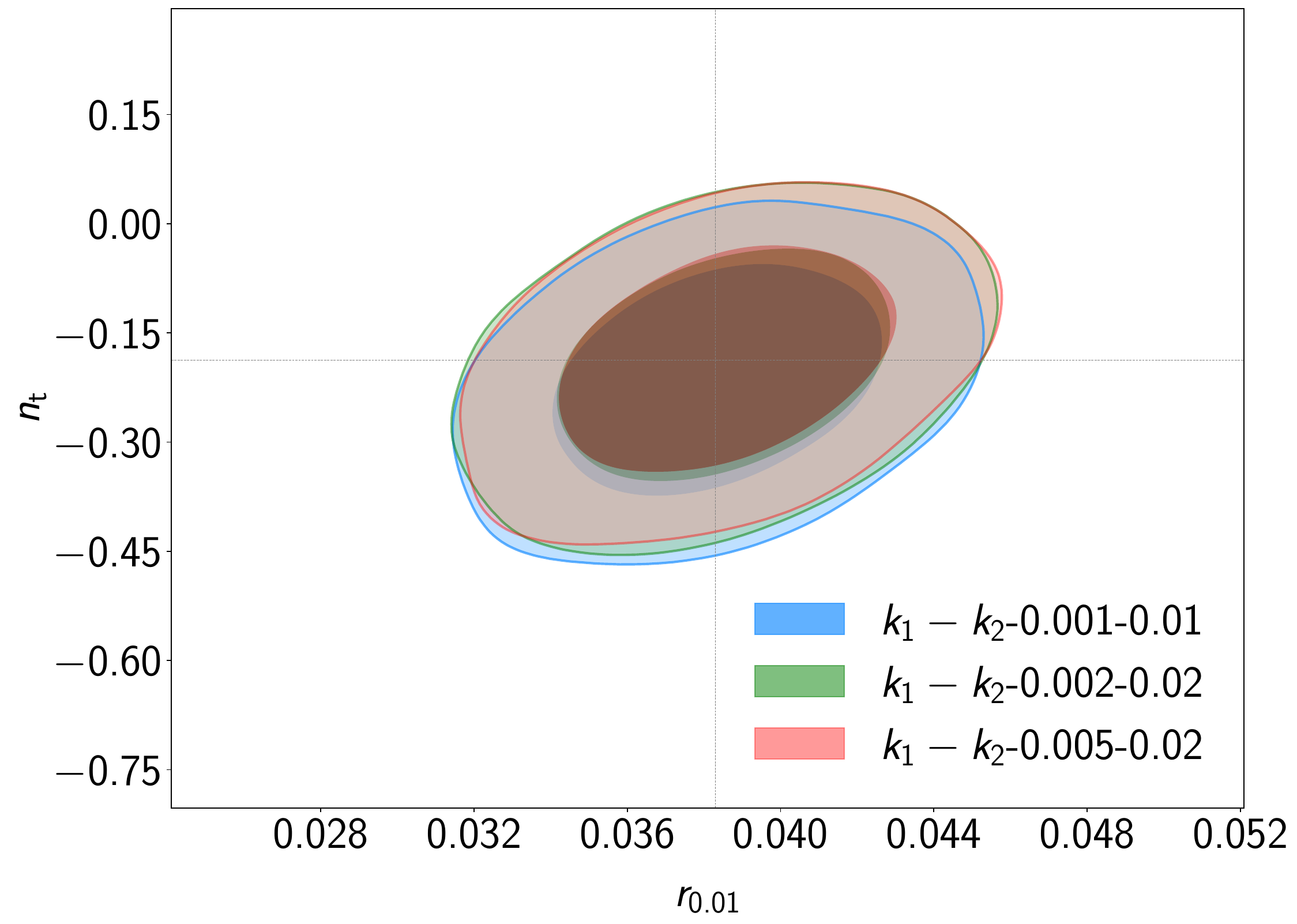}\includegraphics[width=0.33\textwidth]{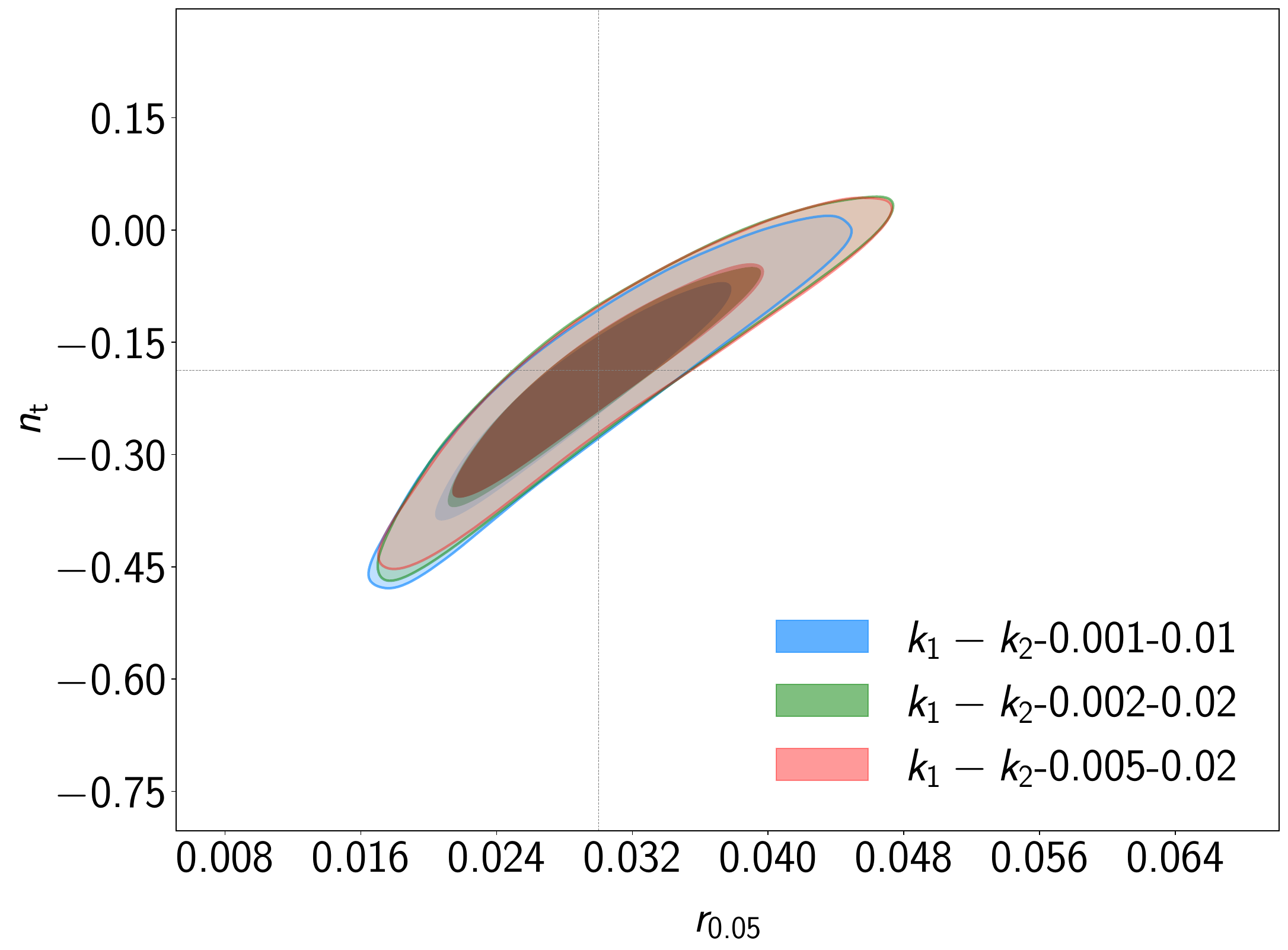}
\caption{Realistic $n_\mathrm{t}-$ case forecasts for the marginalized $(r_1,r_2)$ (left panel), $(r_{0.01},n_{\rm t})$ (middle panel), $(r_{0.05},n_{\rm t})$ (right panel) contours obtained by a two-scale analysis with three different pairs of scales (in blue, green, and red), using simulated data based on fiducial tensor parameters $r_{0.05}=0.030$ and $n_{\rm t}=-0.16$.}
\label{fig:fig:r0p03}
\end{figure*}

\subsubsection{Positive tensor tilt $(n_\mathrm{t}+)$}

For the case of a positive tensor tilt $n_\mathrm{t}=0.3$ we set $r_{0.05}=0.07$ in our fiducial model, again compatible with 95\% CL of \Planck+BK18. We present the posterior constraints for the tensor parameters in Table~\ref{tab:NT07} and in Fig.~\ref{fig:fig:r0p07_1}. 
Again the couple $k_1$---$k_2$=$0.002$---$0.02$ leads to a 
degeneracy between $r_1$ and $r_2$,
but any of the choices would recover the input value of $r_{0.01}$ equally well. The measurement precision is now $\sigma(r_{0.001})\approx0.0058$ and $\sigma(r_{0.01})\approx0.0028$, i.e., slightly better compared to the IC case. 
As for the IC, also in this case we reach $\sigma(n_\mathrm{t})\approx 0.10$.
The measured tensor tilt is clearly positive and differs from zero by more than $2\,\sigma$.

\subsubsection{Modified sound speed $(n_\mathrm{t}-)$}

We conclude with a negative tensor tilt due to a modified sound speed. We use
a value compatible with the current constraints, $c_s=0.02$, and assume $r_{0.05}=0.03$, which gives $n_\mathrm{t}=-0.1875$. We present the constraints for the tensor parameters in Table~\ref{tab:CS03} and in Fig.~\ref{fig:fig:r0p03}.

Also in this case, $k_1$---$k_2$=$0.001$---$0.01$ Mpc$^{-1}$ performs marginally better than $0.005$---$0.02$ Mpc$^{-1}$, whereas $0.002$---$0.02$ Mpc$^{-1}$ leads again to a degeneracy for the primary tensor parameters. The forecasted measurement precision is $\sigma(r_{0.001})\approx0.0014$ and $\sigma(r_{0.01})\approx0.003$. The measured tensor tilt is clearly negative and differs from zero at $2\,\sigma$.

\renewcommand{\dblfloatpagefraction}{0.7}
\subsection{The importance of space-based mission\label{sec:spacebased}}

\begin{figure}[b]
\includegraphics[width=0.465\textwidth]{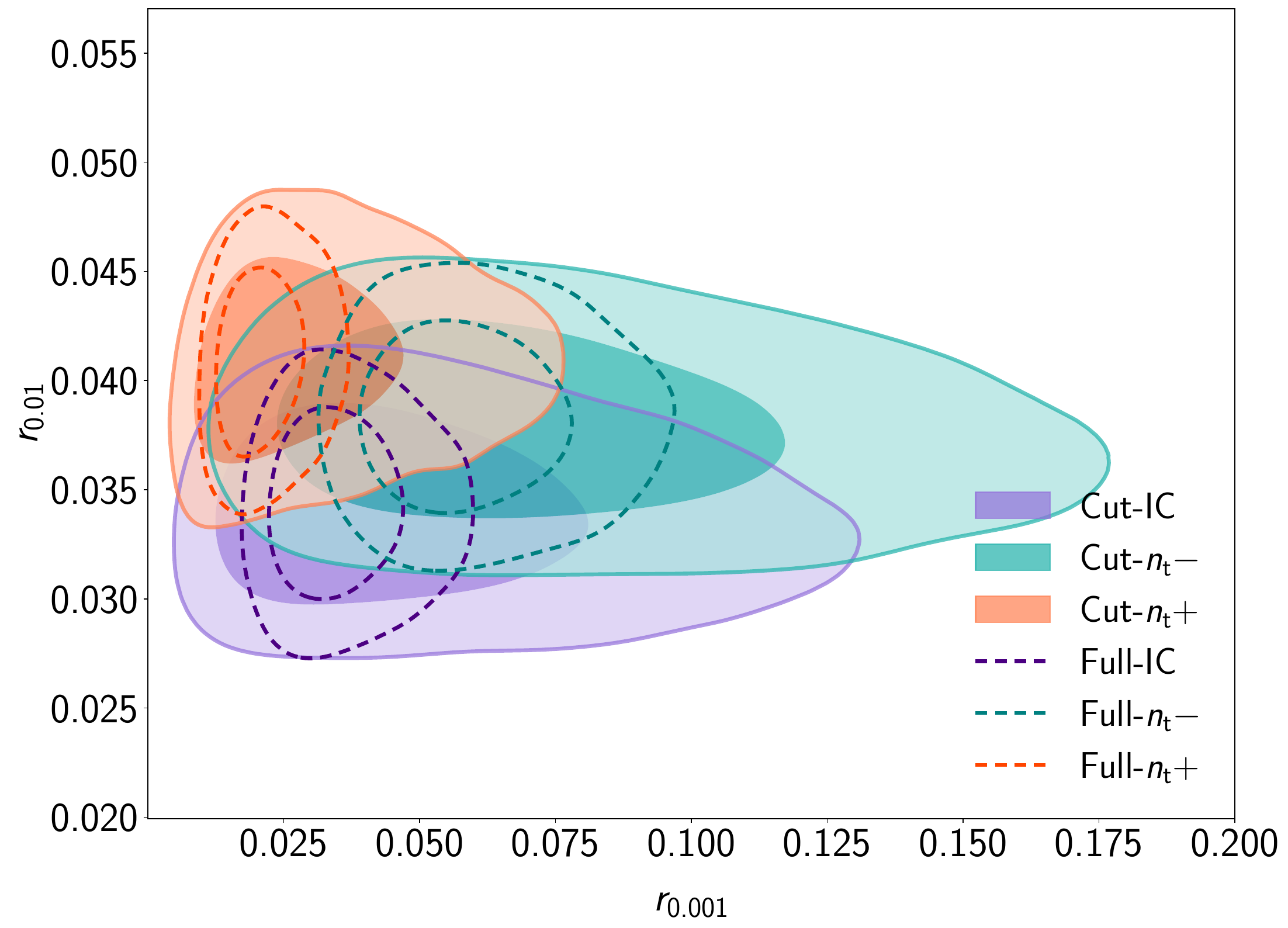}\\
\vspace{-3.3mm}
\caption{Two-dimensional posterior distributions for the three realistic fiducial models with simulated LiteBIRD-like data 
without the low multipoles $\ell \in [2,19]$ (``Cut", shaded regions) and with the low multipoles also included (``Full", dashed lines).
}
\label{fig:LBcut}
\end{figure}

We now study the impact of the low multipoles which are accessible only by observing a sufficiently large fraction of the sky, one of the main advantages of space missions. In order to mimic what could happen with a ground-based instrument, we use the same setup as above but with $\ell=[2,19]$ for the $B$-mode polarization removed.
The results for $k_1$---$k_2$=$0.001$---$0.01$ Mpc$^{-1}$ in Fig.~\ref{fig:LBcut} show how the uncertainties on $r_{0.001}$ more than double for any of the three models. We observe this degradation of constraints in $r_1$ also for the other sets of scales.
Naturally, $r_{0.01}$
(as well as $r_{0.02}$)
stays largely unaffected since $k=0.01\,$Mpc$^{-1}$ corresponds to multipoles larger than 20 (indeed $\ell \sim 70$). Importantly, without the space-based low-multipole data even the 68\% CL posterior regions of our representative three fiducial models overlap, whereas with the low multipoles included the positive tensor tilt is clearly distinguishable from the negative tilt, and the IC case only marginally overlaps with the $n_\mathrm{t}-$ case at 68\% CL.
\begin{table}
\begin{tabular}{ p{1.1cm} p{2.3cm} p{2.3cm} p{2.3cm}  }
 \hline
 \hline
\noalign{\vskip 3pt}
 \multicolumn{4}{c}{Inflation consistency (IC)} \\
\noalign{\hrule\vskip 3pt}
Pars & $0.001-0.01$&$0.002-0.02$&$0.005-0.02$\\
\noalign{\hrule\vskip 3pt}
$r_1$ & $0.035_{-0.010}^{+0.006}$ & $0.040_{-0.007}^{+0.005}$ &$0.034 \pm 0.003$\\
\noalign{\vskip 2pt}
$r_2$ & $0.034 \pm 0.003$ & $0.035_{-0.005}^{+0.004}$ & $0.035_{-0.005}^{+0.004}$\\
\noalign{\vskip 1pt}
\noalign{\hrule\vskip 3pt}
$r_{0.01}$ & $0.034 \pm 0.003$& $0.035\pm 0.003$ &$0.035 \pm 0.003$  \\ 
$n_\mathrm{t}$  & $-0.03\pm 0.10$ & $-0.02\pm 0.10$ &$-0.01 \pm 0.10$ \\
\hline
\hline\\
\noalign{\vskip -15pt}
\end{tabular}
\caption{\label{tab:IC036}68\% CL results for the tensor parameters when jointly fitting the $\Lambda$CDM parameters and $r_1$ and $r_2$ to the simulated LiteBIRD-like data. In this case, the input model 
has $r_{0.05}=0.036$ and $n_\mathrm{t}=-r/8$
(i.e., $r_{0.01}=0.0343$ and $n_\mathrm{t} = -0.0045$).} 
\end{table}
\begin{table}
\begin{tabular}{ p{1.1cm} p{2.3cm} p{2.3cm} p{2.3cm}  }
 \hline
 \hline
 \noalign{\vskip 3pt}
 \multicolumn{4}{c}{Positive tensor tilt ($n_\mathrm{t}+$) } \\
\noalign{\hrule\vskip 3pt}
Pars & $0.001-0.01$&$0.002-0.02$&$0.005-0.02$\\
\noalign{\hrule\vskip 3pt}
$r_1$ & $0.021_{-0.006}^{+0.004}$ & $0.025_{-0.005}^{+0.004}$ &$0.033 \pm 0.003$\\
\noalign{\vskip 2pt}
$r_2$ & $0.041\pm 0.003$ & $0.051 \pm 0.005$ & $0.052_{-0.006}^{+0.005}$ \\
\noalign{\vskip 1pt}
\noalign{\hrule\vskip 3pt}
$r_{0.01}$  & $0.041\pm 0.003$& $0.041 \pm 0.003$ &$0.041 \pm 0.003$  \\ 
$n_\mathrm{t}$  & $0.27\pm 0.10$& $0.28 \pm 0.10$ &$0.30 \pm 0.10$ \\
\hline
\hline\\
\noalign{\vskip -15pt}
\end{tabular}
\caption{\label{tab:NT07} The same as Table~\ref{tab:IC036}, but now simulating the LiteBIRD-like data by using a fiducial model with $r_{0.05}=0.07$ and $n_\mathrm{t}=0.3$ (i.e., $r_{0.01}=0.0408$). 
} 
\end{table}
\begin{table}
\begin{tabular}{ p{1.1cm} p{2.3cm} p{2.3cm} p{2.3cm}  }
 \hline
 \hline
 \noalign{\vskip 3pt}
 \multicolumn{4}{c}{Negative tensor tilt ($n_\mathrm{t}-$)} \\
 \noalign{\hrule\vskip 3pt}
Pars & $0.001-0.01$&$0.002-0.02$&$0.005-0.02$\\
\noalign{\hrule\vskip 3pt}
$r_1$ & $0.059_{-0.016}^{+0.010}$ & $0.050_{-0.009}^{+0.006}$ &$0.043_{-0.004}^{+0.003}$\\
\noalign{\vskip 2pt}
$r_2$ & $0.038\pm 0.003$ & $0.035 \pm 0.004$ & $0.035 \pm 0.004$ \\
\noalign{\vskip 1pt}
\noalign{\hrule\vskip 3pt}
$r_{0.01}$  & $0.038\pm 0.003$& $0.039\pm 0.003$ &$0.039\pm 0.003$  \\ 
$n_\mathrm{t}$  & $-0.22 \pm 0.10$& $-0.19 \pm 0.10$ &$-0.19 \pm 0.10$ \\
\hline
\hline\\
\noalign{\vskip -15pt}
\end{tabular}
\caption{\label{tab:CS03}  The same as Table~\ref{tab:IC036}, but now simulating the LiteBIRD-like data by using a fiducial model with
the modified sound speed, adopting currently allowed values $r_{0.05}=0.03$ and $n_\mathrm{t} = -0.1875$ (giving $r_{0.01}=0.0384$).
} 
\end{table}

\begin{figure}
\includegraphics[width=0.48\textwidth]{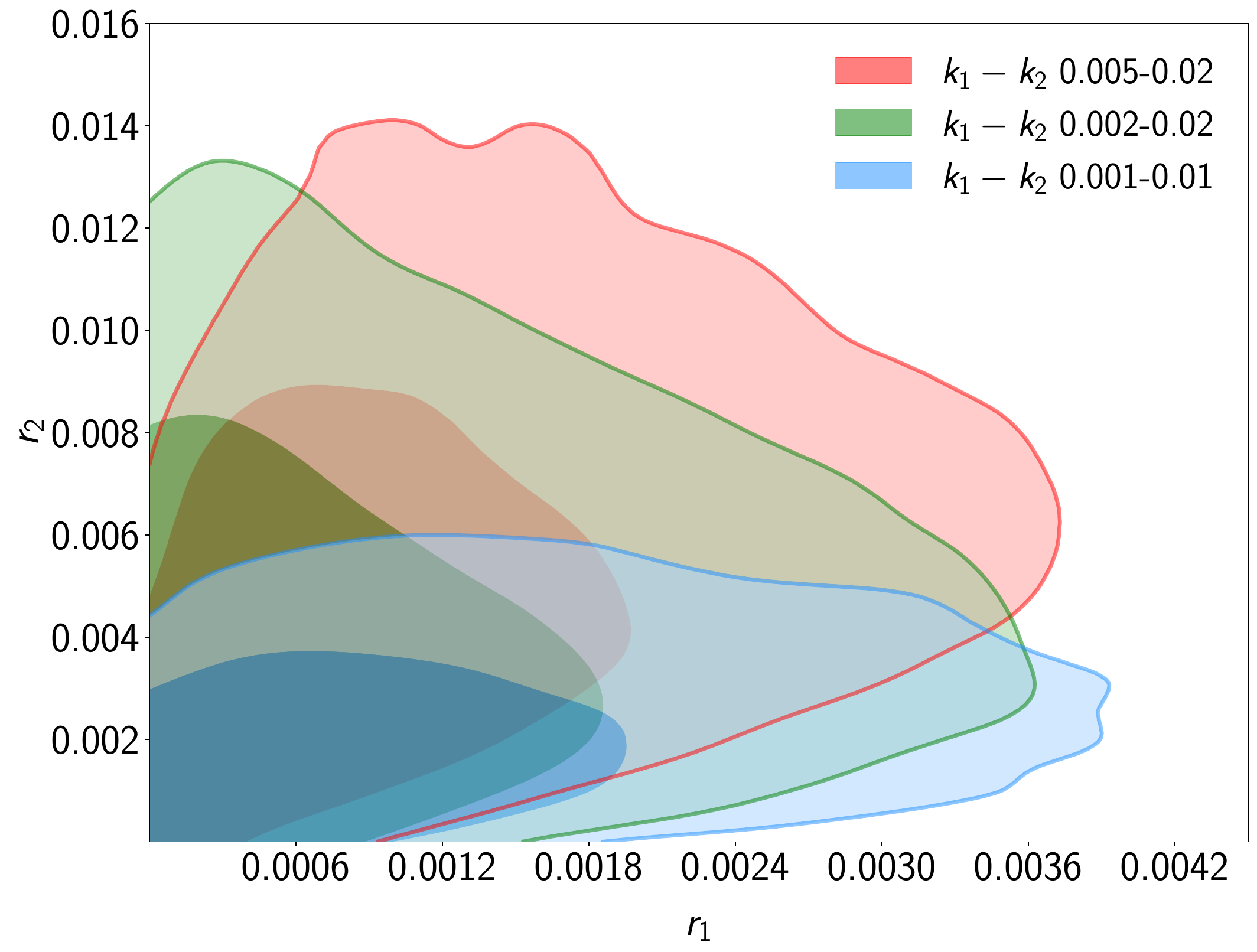}
\caption{Two-dimensional posterior distributions for the case of simulated LiteBIRD-like data with zero tensor contribution in the sky.}
\label{fig:LB0}
\end{figure}
\begin{figure*}
{\includegraphics[width=0.45\textwidth]{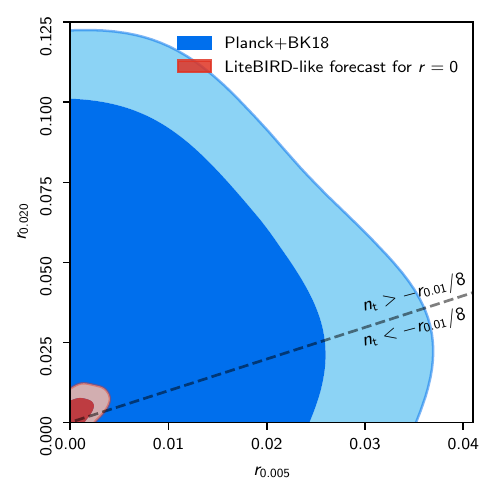}
\includegraphics[width=0.45\textwidth]{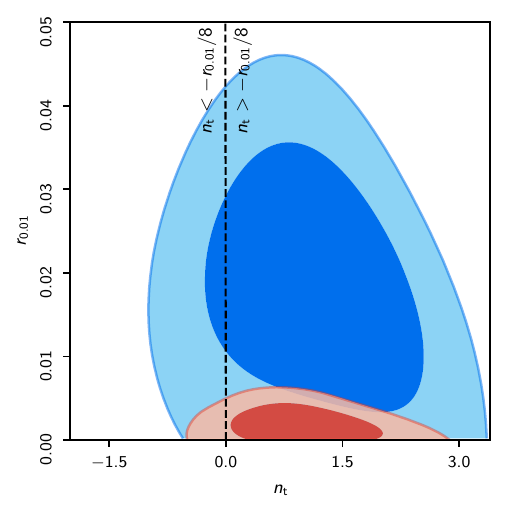}\\
\vspace{-2mm}
\includegraphics[width=0.97\textwidth]{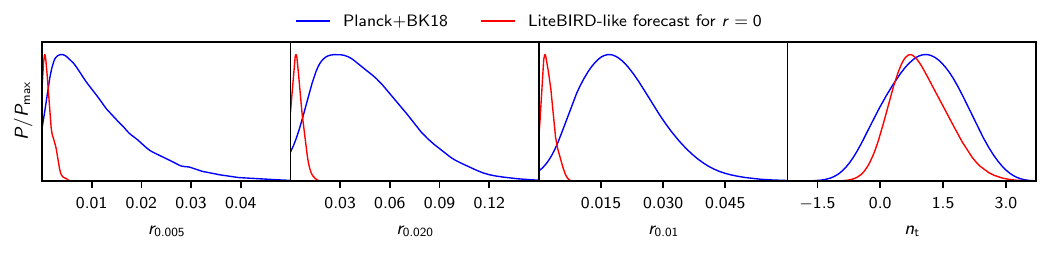}}
\caption{Comparison of expected tensor-parameter constraints from future experiments for a null case together with the current bounds from the \Planck+BK18 data, with the same notation as in Fig.~\ref{fig:realdata}.
The primary sampling parameters were the standard $\Lambda$CDM ones and $(r_{0.005},\,r_{0.02})$ for the tensor contribution (see the first panel). Other parameters shown in the plots are derived ones. 
Red regions and lines represent a forecast with LiteBIRD-like specifications, using a fiducial model with $r=0$, which highlights the huge discovery potential of LiteBIRD.} 
\label{fig:final}
\end{figure*}

\subsection{Null case \label{sec:nullcase}}

In addition to the three cases with $r \ne 0$,
we also test $r=0$, representative of the case that inflation or its alternatives generate gravitational waves with an amplitude below the threshold of detection of future $B$-mode polarization experiments.
The results are shown in Fig.~\ref{fig:LB0} (and in Fig.~\ref{fig:final} where we compare to the constraints given by the real data). We obtain the following 95\% CL upper bounds: $r_{0.001} < 0.0032$ and $r_{0.01} <0.0048$
(or $r_{0.005}<0.0027$
and $r_{0.02} < 0.01$), whereas our result with the current \Planck+BK18 data was $r_{0.005}<0.030$ and $r_{0.02} < 0.098$. Thus, in the possible null case, LiteBIRD would lead to 10 times tighter constraints on both primary tensor parameters than achieved by \Planck+BK18. As seen in the first panel of Fig.~\ref{fig:final}, this means that the area covered by the 95\% CL region in the ($r_{0.005},\,r_{0.02}$) plane shrinks more than by a factor of 100.

Equally impressively, any point outside of the red region in the first panel of Fig.~\ref{fig:final} can be regarded at least as a two-sigma detection zone of a non-zero tensor contribution. As we move to the outer limits of the blue region (the currently allowed 95\% CL region) the detection of a non-zero $r$ by a LiteBIRD-like experiment would be at the $10\,\sigma$ level, as we have seen in the previous subsections.
Let us finally note that in the case of a fiducial with zero tensor contribution, the LiteBIRD-like data lead to only a marginally narrower 95\% CL range for $n_\mathrm{t}$
than the current data, and the posterior peaks again at $n_\mathrm{t} \approx 1$, confirming the explanation of Sec.~\ref{sec:realdata} and showing that the effect of the prior peaking at $n_\mathrm{t} \approx 0$ (see the upper panel of Fig.~\ref{fig:ntprior}) is negligible. 

\section{Conclusions\label{sec:conclusions}}

We have obtained constraints on the amplitude and tilt of the primordial tensor mode by the most recent \Planck\ 
and BICEP/\Keck\ Array 2018 data, employing a two-scale analysis where
the sampling parameters for the tensor power spectrum are (independent) tensor-to-scalar ratios at two different scales. 
This is a minimal extension of the analysis with $n_\mathrm{t}$ fixed with respect to more ambitious reconstructions of the primordial tensor power spectrum \cite{Hiramatsu:2018nfa,Campeti:2019ylm}.

Our 95\% CL
constraints $r_{0.005} < 0.030$ and $r_{0.02} < 0.098$ improve by nearly a factor of 2 those obtained from \Planck\ 2018 data in combination with the previous $B$-mode polarization 
BK15 likelihood in \cite{Planck:2018jri}. The \Planck+BK18 95\% CL constraints on the derived tensor parameters are $r_{0.01} < 0.039$
and $-0.6 < n_{\rm t} < 2.7$.
As in \cite{Planck:2018jri},
we also report the results in combination with the upper bound on the stochastic gravitational wave background at much smaller scales, provided by the LIGO\&Virgo 2016 observing season, which excludes most positive values of the primordial tensor tilt:
$r_{0.01} < 0.039$ and $-0.8 < n_{\rm t} < 0.5$ at 95\% CL.

We have then forecasted how a two-scale analysis performs
with future $B$-mode polarization data.
As a representative experiment for future polarization data, we have considered conservative specifications for a LiteBIRD-like space-based mission. Given its capability to probe both the reionization and recombination peaks in the $B$-mode power spectrum, we had the possibility to study different choices of the two scales and to show how the results depend on this choice. 
 
We have also considered different fiducial values for the primordial tensor power spectrum, including the tensor-to-scalar  consistency condition and two cases with $n_\mathrm{t}$ positive and negative, respectively. 
Whereas distinguishing $n_{\rm t}=-r/8$ from the exact scale invariance is out of reach as expected \cite{Knox:2002pe}, we have shown how with these LiteBIRD-like specifications we could detect at $\sigma(n_{\rm t}) \sim 0.1$, largely independent from any reasonable choice of scales, theoretically motivated departures from $n_\mathrm{t}=-r/8$ consistent with the current bounds. Accessing the low multipoles, virtually doable only by CMB space missions, is essential for reaching these results, as discussed in Sec.~\ref{sec:spacebased} and shown in Fig.~\ref{fig:LBcut}.
We have also shown in Fig.~\ref{fig:final} the huge LiteBIRD-like discovery space compared to the current bounds
when $n_\mathrm{t}$ is allowed to vary.
We conclude reminding that the results presented here are conservative with respect to Ref.~\cite{LiteBIRD:2022cnt} and could be further improved by de-lensing, but show that a space mission, such as LiteBIRD, accessing the low multipoles is the most suitable for characterizing the primordial tensor spectrum with the minimal assumptions.

\section*{Acknowledgments}

DP and FF acknowledge financial support by ASI Grant 2016-24-H.0 and the agreement n. 2020-9-HH.0 ASI-UniRM2 ``Partecipazione italiana alla fase A della missione LiteBIRD". JV acknowledges funding from The Finnish Cultural Foundation (2020--21) and Ruth och Nils-Erik Stenb\"acks stiftelse (2022). We acknowledge the use of the INAF-OAS institute HPC cluster, and we acknowledge the use of the computing centre of Cineca under the agreement INFN-InDark. Part of the analysis was performed using computational resources provided by CSC --- IT Center for Science, Finland.
MH was supported by
the World Premier International Research Center Initiative (WPI) of MEXT,
and by JSPS KAKENHI Grant Number 22H04945.


\bibliography{apssamp}

\begin{thebibliography}{51}%
\makeatletter
\providecommand \@ifxundefined [1]{%
 \@ifx{#1\undefined}
}%
\providecommand \@ifnum [1]{%
 \ifnum #1\expandafter \@firstoftwo
 \else \expandafter \@secondoftwo
 \fi
}%
\providecommand \@ifx [1]{%
 \ifx #1\expandafter \@firstoftwo
 \else \expandafter \@secondoftwo
 \fi
}%
\providecommand \natexlab [1]{#1}%
\providecommand \enquote  [1]{``#1''}%
\providecommand \bibnamefont  [1]{#1}%
\providecommand \bibfnamefont [1]{#1}%
\providecommand \citenamefont [1]{#1}%
\providecommand \href@noop [0]{\@secondoftwo}%
\providecommand \href [0]{\begingroup \@sanitize@url \@href}%
\providecommand \@href[1]{\@@startlink{#1}\@@href}%
\providecommand \@@href[1]{\endgroup#1\@@endlink}%
\providecommand \@sanitize@url [0]{\catcode `\\12\catcode `\$12\catcode
  `\&12\catcode `\#12\catcode `\^12\catcode `\_12\catcode `\%12\relax}%
\providecommand \@@startlink[1]{}%
\providecommand \@@endlink[0]{}%
\providecommand \url  [0]{\begingroup\@sanitize@url \@url }%
\providecommand \@url [1]{\endgroup\@href {#1}{\urlprefix }}%
\providecommand \urlprefix  [0]{URL }%
\providecommand \Eprint [0]{\href }%
\providecommand \doibase [0]{https://doi.org/}%
\providecommand \selectlanguage [0]{\@gobble}%
\providecommand \bibinfo  [0]{\@secondoftwo}%
\providecommand \bibfield  [0]{\@secondoftwo}%
\providecommand \translation [1]{[#1]}%
\providecommand \BibitemOpen [0]{}%
\providecommand \bibitemStop [0]{}%
\providecommand \bibitemNoStop [0]{.\EOS\space}%
\providecommand \EOS [0]{\spacefactor3000\relax}%
\providecommand \BibitemShut  [1]{\csname bibitem#1\endcsname}%
\let\auto@bib@innerbib\@empty
\bibitem [{\citenamefont {Starobinsky}(1979)}]{Starobinsky:1979ty}%
  \BibitemOpen
  \bibfield  {author} {\bibinfo {author} {\bibfnamefont {A.~A.}\ \bibnamefont
  {Starobinsky}},\ }\bibfield  {title} {\bibinfo {title} {{Spectrum of relict
  gravitational radiation and the early state of the universe}},\ }\href@noop
  {} {\bibfield  {journal} {\bibinfo  {journal} {JETP Lett.}\ }\textbf
  {\bibinfo {volume} {30}},\ \bibinfo {pages} {682} (\bibinfo {year}
  {1979})}\BibitemShut {NoStop}%
\bibitem [{\citenamefont {Kamionkowski}\ \emph {et~al.}(1997)\citenamefont
  {Kamionkowski}, \citenamefont {Kosowsky},\ and\ \citenamefont
  {Stebbins}}]{Kamionkowski:1996zd}%
  \BibitemOpen
  \bibfield  {author} {\bibinfo {author} {\bibfnamefont {M.}~\bibnamefont
  {Kamionkowski}}, \bibinfo {author} {\bibfnamefont {A.}~\bibnamefont
  {Kosowsky}},\ and\ \bibinfo {author} {\bibfnamefont {A.}~\bibnamefont
  {Stebbins}},\ }\bibfield  {title} {\bibinfo {title} {{A Probe of primordial
  gravity waves and vorticity}},\ }\href
  {https://doi.org/10.1103/PhysRevLett.78.2058} {\bibfield  {journal} {\bibinfo
   {journal} {Phys. Rev. Lett.}\ }\textbf {\bibinfo {volume} {78}},\ \bibinfo
  {pages} {2058} (\bibinfo {year} {1997})},\ \Eprint
  {https://arxiv.org/abs/astro-ph/9609132} {arXiv:astro-ph/9609132}
  \BibitemShut {NoStop}%
\bibitem [{\citenamefont {Seljak}\ and\ \citenamefont
  {Zaldarriaga}(1997)}]{Seljak:1996gy}%
  \BibitemOpen
  \bibfield  {author} {\bibinfo {author} {\bibfnamefont {U.}~\bibnamefont
  {Seljak}}\ and\ \bibinfo {author} {\bibfnamefont {M.}~\bibnamefont
  {Zaldarriaga}},\ }\bibfield  {title} {\bibinfo {title} {{Signature of gravity
  waves in polarization of the microwave background}},\ }\href
  {https://doi.org/10.1103/PhysRevLett.78.2054} {\bibfield  {journal} {\bibinfo
   {journal} {Phys. Rev. Lett.}\ }\textbf {\bibinfo {volume} {78}},\ \bibinfo
  {pages} {2054} (\bibinfo {year} {1997})},\ \Eprint
  {https://arxiv.org/abs/astro-ph/9609169} {arXiv:astro-ph/9609169}
  \BibitemShut {NoStop}%
\bibitem [{\citenamefont {Ade}\ \emph {et~al.}(2015)\citenamefont {Ade} \emph
  {et~al.}}]{BICEP2:2015nss}%
  \BibitemOpen
  \bibfield  {author} {\bibinfo {author} {\bibfnamefont {P.~A.~R.}\
  \bibnamefont {Ade}} \emph {et~al.} (\bibinfo {collaboration} {BICEP2,
  Planck}),\ }\bibfield  {title} {\bibinfo {title} {{Joint Analysis of
  BICEP2/{\it{Keck}} Array and {\it{Planck}} Data}},\ }\href
  {https://doi.org/10.1103/PhysRevLett.114.101301} {\bibfield  {journal}
  {\bibinfo  {journal} {Phys. Rev. Lett.}\ }\textbf {\bibinfo {volume} {114}},\
  \bibinfo {pages} {101301} (\bibinfo {year} {2015})},\ \Eprint
  {https://arxiv.org/abs/1502.00612} {arXiv:1502.00612 [astro-ph.CO]}
  \BibitemShut {NoStop}%
\bibitem [{\citenamefont {Ade}\ \emph {et~al.}(2016{\natexlab{a}})\citenamefont
  {Ade} \emph {et~al.}}]{BICEP2:2015xme}%
  \BibitemOpen
  \bibfield  {author} {\bibinfo {author} {\bibfnamefont {P.~A.~R.}\
  \bibnamefont {Ade}} \emph {et~al.} (\bibinfo {collaboration} {BICEP2, Keck
  Array}),\ }\bibfield  {title} {\bibinfo {title} {{Improved Constraints on
  Cosmology and Foregrounds from BICEP2 and Keck Array Cosmic Microwave
  Background Data with Inclusion of 95 GHz Band}},\ }\href
  {https://doi.org/10.1103/PhysRevLett.116.031302} {\bibfield  {journal}
  {\bibinfo  {journal} {Phys. Rev. Lett.}\ }\textbf {\bibinfo {volume} {116}},\
  \bibinfo {pages} {031302} (\bibinfo {year} {2016}{\natexlab{a}})},\ \Eprint
  {https://arxiv.org/abs/1510.09217} {arXiv:1510.09217 [astro-ph.CO]}
  \BibitemShut {NoStop}%
\bibitem [{\citenamefont {Ade}\ \emph {et~al.}(2018)\citenamefont {Ade} \emph
  {et~al.}}]{BICEP2:2018kqh}%
  \BibitemOpen
  \bibfield  {author} {\bibinfo {author} {\bibfnamefont {P.~A.~R.}\
  \bibnamefont {Ade}} \emph {et~al.} (\bibinfo {collaboration} {BICEP2, Keck
  Array}),\ }\bibfield  {title} {\bibinfo {title} {{BICEP2 / Keck Array x:
  Constraints on Primordial Gravitational Waves using Planck, WMAP, and New
  BICEP2/Keck Observations through the 2015 Season}},\ }\href
  {https://doi.org/10.1103/PhysRevLett.121.221301} {\bibfield  {journal}
  {\bibinfo  {journal} {Phys. Rev. Lett.}\ }\textbf {\bibinfo {volume} {121}},\
  \bibinfo {pages} {221301} (\bibinfo {year} {2018})},\ \Eprint
  {https://arxiv.org/abs/1810.05216} {arXiv:1810.05216 [astro-ph.CO]}
  \BibitemShut {NoStop}%
\bibitem [{\citenamefont {Ade}\ \emph {et~al.}(2021)\citenamefont {Ade} \emph
  {et~al.}}]{BICEP:2021xfz}%
  \BibitemOpen
  \bibfield  {author} {\bibinfo {author} {\bibfnamefont {P.~A.~R.}\
  \bibnamefont {Ade}} \emph {et~al.} (\bibinfo {collaboration} {BICEP, Keck}),\
  }\bibfield  {title} {\bibinfo {title} {{Improved Constraints on Primordial
  Gravitational Waves using Planck, WMAP, and BICEP/Keck Observations through
  the 2018 Observing Season}},\ }\href
  {https://doi.org/10.1103/PhysRevLett.127.151301} {\bibfield  {journal}
  {\bibinfo  {journal} {Phys. Rev. Lett.}\ }\textbf {\bibinfo {volume} {127}},\
  \bibinfo {pages} {151301} (\bibinfo {year} {2021})},\ \Eprint
  {https://arxiv.org/abs/2110.00483} {arXiv:2110.00483 [astro-ph.CO]}
  \BibitemShut {NoStop}%
\bibitem [{Note1()}]{Note1}%
  \BibitemOpen
  \bibinfo {note} {We denote the tensor-to-scalar ratio at this scale simply by
  $r$ or occasionally by $r_{0.05}$ and at any other scale by adding a
  subscript indicating the corresponding wavenumber. Integer subscripts $1$ and
  $2$ refer to the scales of our two-scale parametrization, explained in
  Sec.~\ref {sec:two-scale}.}\BibitemShut {Stop}%
\bibitem [{\citenamefont {Akrami}\ \emph
  {et~al.}(2020{\natexlab{a}})\citenamefont {Akrami} \emph
  {et~al.}}]{Planck:2018jri}%
  \BibitemOpen
  \bibfield  {author} {\bibinfo {author} {\bibfnamefont {Y.}~\bibnamefont
  {Akrami}} \emph {et~al.} (\bibinfo {collaboration} {Planck}),\ }\bibfield
  {title} {\bibinfo {title} {{Planck 2018 results. X. Constraints on
  inflation}},\ }\href {https://doi.org/10.1051/0004-6361/201833887} {\bibfield
   {journal} {\bibinfo  {journal} {Astron. Astrophys.}\ }\textbf {\bibinfo
  {volume} {641}},\ \bibinfo {pages} {A10} (\bibinfo {year}
  {2020}{\natexlab{a}})},\ \Eprint {https://arxiv.org/abs/1807.06211}
  {arXiv:1807.06211 [astro-ph.CO]} \BibitemShut {NoStop}%
\bibitem [{\citenamefont {Ade}\ \emph {et~al.}(2016{\natexlab{b}})\citenamefont
  {Ade} \emph {et~al.}}]{Planck:2015sxf}%
  \BibitemOpen
  \bibfield  {author} {\bibinfo {author} {\bibfnamefont {P.~A.~R.}\
  \bibnamefont {Ade}} \emph {et~al.} (\bibinfo {collaboration} {Planck}),\
  }\bibfield  {title} {\bibinfo {title} {{Planck 2015 results. XX. Constraints
  on inflation}},\ }\href {https://doi.org/10.1051/0004-6361/201525898}
  {\bibfield  {journal} {\bibinfo  {journal} {Astron. Astrophys.}\ }\textbf
  {\bibinfo {volume} {594}},\ \bibinfo {pages} {A20} (\bibinfo {year}
  {2016}{\natexlab{b}})},\ \Eprint {https://arxiv.org/abs/1502.02114}
  {arXiv:1502.02114 [astro-ph.CO]} \BibitemShut {NoStop}%
\bibitem [{\citenamefont {Garriga}\ and\ \citenamefont
  {Mukhanov}(1999)}]{Garriga:1999vw}%
  \BibitemOpen
  \bibfield  {author} {\bibinfo {author} {\bibfnamefont {J.}~\bibnamefont
  {Garriga}}\ and\ \bibinfo {author} {\bibfnamefont {V.~F.}\ \bibnamefont
  {Mukhanov}},\ }\bibfield  {title} {\bibinfo {title} {{Perturbations in
  k-inflation}},\ }\href {https://doi.org/10.1016/S0370-2693(99)00602-4}
  {\bibfield  {journal} {\bibinfo  {journal} {Phys. Lett. B}\ }\textbf
  {\bibinfo {volume} {458}},\ \bibinfo {pages} {219} (\bibinfo {year}
  {1999})},\ \Eprint {https://arxiv.org/abs/hep-th/9904176}
  {arXiv:hep-th/9904176} \BibitemShut {NoStop}%
\bibitem [{\citenamefont {Kobayashi}\ \emph {et~al.}(2010)\citenamefont
  {Kobayashi}, \citenamefont {Yamaguchi},\ and\ \citenamefont
  {Yokoyama}}]{Kobayashi:2010cm}%
  \BibitemOpen
  \bibfield  {author} {\bibinfo {author} {\bibfnamefont {T.}~\bibnamefont
  {Kobayashi}}, \bibinfo {author} {\bibfnamefont {M.}~\bibnamefont
  {Yamaguchi}},\ and\ \bibinfo {author} {\bibfnamefont {J.}~\bibnamefont
  {Yokoyama}},\ }\bibfield  {title} {\bibinfo {title} {{G-inflation: Inflation
  driven by the Galileon field}},\ }\href
  {https://doi.org/10.1103/PhysRevLett.105.231302} {\bibfield  {journal}
  {\bibinfo  {journal} {Phys. Rev. Lett.}\ }\textbf {\bibinfo {volume} {105}},\
  \bibinfo {pages} {231302} (\bibinfo {year} {2010})},\ \Eprint
  {https://arxiv.org/abs/1008.0603} {arXiv:1008.0603 [hep-th]} \BibitemShut
  {NoStop}%
\bibitem [{\citenamefont {Ashoorioon}\ \emph {et~al.}(2014)\citenamefont
  {Ashoorioon}, \citenamefont {Dimopoulos}, \citenamefont {Sheikh-Jabbari},\
  and\ \citenamefont {Shiu}}]{Ashoorioon:2013eia}%
  \BibitemOpen
  \bibfield  {author} {\bibinfo {author} {\bibfnamefont {A.}~\bibnamefont
  {Ashoorioon}}, \bibinfo {author} {\bibfnamefont {K.}~\bibnamefont
  {Dimopoulos}}, \bibinfo {author} {\bibfnamefont {M.~M.}\ \bibnamefont
  {Sheikh-Jabbari}},\ and\ \bibinfo {author} {\bibfnamefont {G.}~\bibnamefont
  {Shiu}},\ }\bibfield  {title} {\bibinfo {title} {{Reconciliation of High
  Energy Scale Models of Inflation with Planck}},\ }\href
  {https://doi.org/10.1088/1475-7516/2014/02/025} {\bibfield  {journal}
  {\bibinfo  {journal} {JCAP}\ }\textbf {\bibinfo {volume} {02}},\ \bibinfo
  {pages} {025}},\ \Eprint {https://arxiv.org/abs/1306.4914} {arXiv:1306.4914
  [hep-th]} \BibitemShut {NoStop}%
\bibitem [{\citenamefont {Bartolo}\ \emph {et~al.}(2001)\citenamefont
  {Bartolo}, \citenamefont {Matarrese},\ and\ \citenamefont
  {Riotto}}]{Bartolo:2001rt}%
  \BibitemOpen
  \bibfield  {author} {\bibinfo {author} {\bibfnamefont {N.}~\bibnamefont
  {Bartolo}}, \bibinfo {author} {\bibfnamefont {S.}~\bibnamefont {Matarrese}},\
  and\ \bibinfo {author} {\bibfnamefont {A.}~\bibnamefont {Riotto}},\
  }\bibfield  {title} {\bibinfo {title} {{Adiabatic and isocurvature
  perturbations from inflation: Power spectra and consistency relations}},\
  }\href {https://doi.org/10.1103/PhysRevD.64.123504} {\bibfield  {journal}
  {\bibinfo  {journal} {Phys. Rev. D}\ }\textbf {\bibinfo {volume} {64}},\
  \bibinfo {pages} {123504} (\bibinfo {year} {2001})},\ \Eprint
  {https://arxiv.org/abs/astro-ph/0107502} {arXiv:astro-ph/0107502}
  \BibitemShut {NoStop}%
\bibitem [{\citenamefont {Wands}\ \emph {et~al.}(2002)\citenamefont {Wands},
  \citenamefont {Bartolo}, \citenamefont {Matarrese},\ and\ \citenamefont
  {Riotto}}]{Wands:2002bn}%
  \BibitemOpen
  \bibfield  {author} {\bibinfo {author} {\bibfnamefont {D.}~\bibnamefont
  {Wands}}, \bibinfo {author} {\bibfnamefont {N.}~\bibnamefont {Bartolo}},
  \bibinfo {author} {\bibfnamefont {S.}~\bibnamefont {Matarrese}},\ and\
  \bibinfo {author} {\bibfnamefont {A.}~\bibnamefont {Riotto}},\ }\bibfield
  {title} {\bibinfo {title} {{An Observational test of two-field inflation}},\
  }\href {https://doi.org/10.1103/PhysRevD.66.043520} {\bibfield  {journal}
  {\bibinfo  {journal} {Phys. Rev. D}\ }\textbf {\bibinfo {volume} {66}},\
  \bibinfo {pages} {043520} (\bibinfo {year} {2002})},\ \Eprint
  {https://arxiv.org/abs/astro-ph/0205253} {arXiv:astro-ph/0205253}
  \BibitemShut {NoStop}%
\bibitem [{\citenamefont {Byrnes}\ and\ \citenamefont
  {Wands}(2006)}]{Byrnes:2006fr}%
  \BibitemOpen
  \bibfield  {author} {\bibinfo {author} {\bibfnamefont {C.~T.}\ \bibnamefont
  {Byrnes}}\ and\ \bibinfo {author} {\bibfnamefont {D.}~\bibnamefont {Wands}},\
  }\bibfield  {title} {\bibinfo {title} {{Curvature and isocurvature
  perturbations from two-field inflation in a slow-roll expansion}},\ }\href
  {https://doi.org/10.1103/PhysRevD.74.043529} {\bibfield  {journal} {\bibinfo
  {journal} {Phys. Rev. D}\ }\textbf {\bibinfo {volume} {74}},\ \bibinfo
  {pages} {043529} (\bibinfo {year} {2006})},\ \Eprint
  {https://arxiv.org/abs/astro-ph/0605679} {arXiv:astro-ph/0605679}
  \BibitemShut {NoStop}%
\bibitem [{\citenamefont {Di~Marco}\ and\ \citenamefont
  {Finelli}(2005)}]{DiMarco:2005nq}%
  \BibitemOpen
  \bibfield  {author} {\bibinfo {author} {\bibfnamefont {F.}~\bibnamefont
  {Di~Marco}}\ and\ \bibinfo {author} {\bibfnamefont {F.}~\bibnamefont
  {Finelli}},\ }\bibfield  {title} {\bibinfo {title} {{Slow-roll inflation for
  generalized two-field Lagrangians}},\ }\href
  {https://doi.org/10.1103/PhysRevD.71.123502} {\bibfield  {journal} {\bibinfo
  {journal} {Phys. Rev. D}\ }\textbf {\bibinfo {volume} {71}},\ \bibinfo
  {pages} {123502} (\bibinfo {year} {2005})},\ \Eprint
  {https://arxiv.org/abs/astro-ph/0505198} {arXiv:astro-ph/0505198}
  \BibitemShut {NoStop}%
\bibitem [{\citenamefont {Achucarro}\ \emph {et~al.}(2012)\citenamefont
  {Achucarro}, \citenamefont {Atal}, \citenamefont {Cespedes}, \citenamefont
  {Gong}, \citenamefont {Palma},\ and\ \citenamefont
  {Patil}}]{Achucarro:2012yr}%
  \BibitemOpen
  \bibfield  {author} {\bibinfo {author} {\bibfnamefont {A.}~\bibnamefont
  {Achucarro}}, \bibinfo {author} {\bibfnamefont {V.}~\bibnamefont {Atal}},
  \bibinfo {author} {\bibfnamefont {S.}~\bibnamefont {Cespedes}}, \bibinfo
  {author} {\bibfnamefont {J.-O.}\ \bibnamefont {Gong}}, \bibinfo {author}
  {\bibfnamefont {G.~A.}\ \bibnamefont {Palma}},\ and\ \bibinfo {author}
  {\bibfnamefont {S.~P.}\ \bibnamefont {Patil}},\ }\bibfield  {title} {\bibinfo
  {title} {{Heavy fields, reduced speeds of sound and decoupling during
  inflation}},\ }\href {https://doi.org/10.1103/PhysRevD.86.121301} {\bibfield
  {journal} {\bibinfo  {journal} {Phys. Rev. D}\ }\textbf {\bibinfo {volume}
  {86}},\ \bibinfo {pages} {121301} (\bibinfo {year} {2012})},\ \Eprint
  {https://arxiv.org/abs/1205.0710} {arXiv:1205.0710 [hep-th]} \BibitemShut
  {NoStop}%
\bibitem [{\citenamefont {Maleknejad}\ and\ \citenamefont
  {Sheikh-Jabbari}(2013)}]{Maleknejad:2011jw}%
  \BibitemOpen
  \bibfield  {author} {\bibinfo {author} {\bibfnamefont {A.}~\bibnamefont
  {Maleknejad}}\ and\ \bibinfo {author} {\bibfnamefont {M.~M.}\ \bibnamefont
  {Sheikh-Jabbari}},\ }\bibfield  {title} {\bibinfo {title} {{Gauge-flation:
  Inflation From Non-Abelian Gauge Fields}},\ }\href
  {https://doi.org/10.1016/j.physletb.2013.05.001} {\bibfield  {journal}
  {\bibinfo  {journal} {Phys. Lett. B}\ }\textbf {\bibinfo {volume} {723}},\
  \bibinfo {pages} {224} (\bibinfo {year} {2013})},\ \Eprint
  {https://arxiv.org/abs/1102.1513} {arXiv:1102.1513 [hep-ph]} \BibitemShut
  {NoStop}%
\bibitem [{\citenamefont {Cook}\ and\ \citenamefont
  {Sorbo}(2012)}]{Cook:2011hg}%
  \BibitemOpen
  \bibfield  {author} {\bibinfo {author} {\bibfnamefont {J.~L.}\ \bibnamefont
  {Cook}}\ and\ \bibinfo {author} {\bibfnamefont {L.}~\bibnamefont {Sorbo}},\
  }\bibfield  {title} {\bibinfo {title} {{Particle production during inflation
  and gravitational waves detectable by ground-based interferometers}},\ }\href
  {https://doi.org/10.1103/PhysRevD.85.023534} {\bibfield  {journal} {\bibinfo
  {journal} {Phys. Rev. D}\ }\textbf {\bibinfo {volume} {85}},\ \bibinfo
  {pages} {023534} (\bibinfo {year} {2012})},\ \bibinfo {note} {[Erratum:
  Phys.Rev.D 86, 069901 (2012)]},\ \Eprint {https://arxiv.org/abs/1109.0022}
  {arXiv:1109.0022 [astro-ph.CO]} \BibitemShut {NoStop}%
\bibitem [{\citenamefont {Dimastrogiovanni}\ \emph {et~al.}(2017)\citenamefont
  {Dimastrogiovanni}, \citenamefont {Fasiello},\ and\ \citenamefont
  {Fujita}}]{Dimastrogiovanni:2016fuu}%
  \BibitemOpen
  \bibfield  {author} {\bibinfo {author} {\bibfnamefont {E.}~\bibnamefont
  {Dimastrogiovanni}}, \bibinfo {author} {\bibfnamefont {M.}~\bibnamefont
  {Fasiello}},\ and\ \bibinfo {author} {\bibfnamefont {T.}~\bibnamefont
  {Fujita}},\ }\bibfield  {title} {\bibinfo {title} {{Primordial Gravitational
  Waves from Axion-Gauge Fields Dynamics}},\ }\href
  {https://doi.org/10.1088/1475-7516/2017/01/019} {\bibfield  {journal}
  {\bibinfo  {journal} {JCAP}\ }\textbf {\bibinfo {volume} {01}},\ \bibinfo
  {pages} {019}},\ \Eprint {https://arxiv.org/abs/1608.04216} {arXiv:1608.04216
  [astro-ph.CO]} \BibitemShut {NoStop}%
\bibitem [{\citenamefont {Agrawal}\ \emph {et~al.}(2018)\citenamefont
  {Agrawal}, \citenamefont {Fujita},\ and\ \citenamefont
  {Komatsu}}]{Agrawal:2017awz}%
  \BibitemOpen
  \bibfield  {author} {\bibinfo {author} {\bibfnamefont {A.}~\bibnamefont
  {Agrawal}}, \bibinfo {author} {\bibfnamefont {T.}~\bibnamefont {Fujita}},\
  and\ \bibinfo {author} {\bibfnamefont {E.}~\bibnamefont {Komatsu}},\
  }\bibfield  {title} {\bibinfo {title} {{Large tensor non-Gaussianity from
  axion-gauge field dynamics}},\ }\href
  {https://doi.org/10.1103/PhysRevD.97.103526} {\bibfield  {journal} {\bibinfo
  {journal} {Phys. Rev. D}\ }\textbf {\bibinfo {volume} {97}},\ \bibinfo
  {pages} {103526} (\bibinfo {year} {2018})},\ \Eprint
  {https://arxiv.org/abs/1707.03023} {arXiv:1707.03023 [astro-ph.CO]}
  \BibitemShut {NoStop}%
\bibitem [{\citenamefont {Gasperini}\ and\ \citenamefont
  {Veneziano}(1993)}]{Gasperini:1992em}%
  \BibitemOpen
  \bibfield  {author} {\bibinfo {author} {\bibfnamefont {M.}~\bibnamefont
  {Gasperini}}\ and\ \bibinfo {author} {\bibfnamefont {G.}~\bibnamefont
  {Veneziano}},\ }\bibfield  {title} {\bibinfo {title} {{Pre - big bang in
  string cosmology}},\ }\href {https://doi.org/10.1016/0927-6505(93)90017-8}
  {\bibfield  {journal} {\bibinfo  {journal} {Astropart. Phys.}\ }\textbf
  {\bibinfo {volume} {1}},\ \bibinfo {pages} {317} (\bibinfo {year} {1993})},\
  \Eprint {https://arxiv.org/abs/hep-th/9211021} {arXiv:hep-th/9211021}
  \BibitemShut {NoStop}%
\bibitem [{\citenamefont {Boyle}\ \emph {et~al.}(2004)\citenamefont {Boyle},
  \citenamefont {Steinhardt},\ and\ \citenamefont {Turok}}]{Boyle:2003km}%
  \BibitemOpen
  \bibfield  {author} {\bibinfo {author} {\bibfnamefont {L.~A.}\ \bibnamefont
  {Boyle}}, \bibinfo {author} {\bibfnamefont {P.~J.}\ \bibnamefont
  {Steinhardt}},\ and\ \bibinfo {author} {\bibfnamefont {N.}~\bibnamefont
  {Turok}},\ }\bibfield  {title} {\bibinfo {title} {{The Cosmic gravitational
  wave background in a cyclic universe}},\ }\href
  {https://doi.org/10.1103/PhysRevD.69.127302} {\bibfield  {journal} {\bibinfo
  {journal} {Phys. Rev. D}\ }\textbf {\bibinfo {volume} {69}},\ \bibinfo
  {pages} {127302} (\bibinfo {year} {2004})},\ \Eprint
  {https://arxiv.org/abs/hep-th/0307170} {arXiv:hep-th/0307170} \BibitemShut
  {NoStop}%
\bibitem [{\citenamefont {Brandenberger}\ \emph {et~al.}(2007)\citenamefont
  {Brandenberger}, \citenamefont {Nayeri}, \citenamefont {Patil},\ and\
  \citenamefont {Vafa}}]{Brandenberger:2006xi}%
  \BibitemOpen
  \bibfield  {author} {\bibinfo {author} {\bibfnamefont {R.~H.}\ \bibnamefont
  {Brandenberger}}, \bibinfo {author} {\bibfnamefont {A.}~\bibnamefont
  {Nayeri}}, \bibinfo {author} {\bibfnamefont {S.~P.}\ \bibnamefont {Patil}},\
  and\ \bibinfo {author} {\bibfnamefont {C.}~\bibnamefont {Vafa}},\ }\bibfield
  {title} {\bibinfo {title} {{Tensor Modes from a Primordial Hagedorn Phase of
  String Cosmology}},\ }\href {https://doi.org/10.1103/PhysRevLett.98.231302}
  {\bibfield  {journal} {\bibinfo  {journal} {Phys. Rev. Lett.}\ }\textbf
  {\bibinfo {volume} {98}},\ \bibinfo {pages} {231302} (\bibinfo {year}
  {2007})},\ \Eprint {https://arxiv.org/abs/hep-th/0604126}
  {arXiv:hep-th/0604126} \BibitemShut {NoStop}%
\bibitem [{\citenamefont {Finelli}\ \emph {et~al.}(2018)\citenamefont {Finelli}
  \emph {et~al.}}]{CORE:2016ymi}%
  \BibitemOpen
  \bibfield  {author} {\bibinfo {author} {\bibfnamefont {F.}~\bibnamefont
  {Finelli}} \emph {et~al.} (\bibinfo {collaboration} {CORE}),\ }\bibfield
  {title} {\bibinfo {title} {{Exploring cosmic origins with CORE: Inflation}},\
  }\href {https://doi.org/10.1088/1475-7516/2018/04/016} {\bibfield  {journal}
  {\bibinfo  {journal} {JCAP}\ }\textbf {\bibinfo {volume} {04}},\ \bibinfo
  {pages} {016}},\ \Eprint {https://arxiv.org/abs/1612.08270} {arXiv:1612.08270
  [astro-ph.CO]} \BibitemShut {NoStop}%
\bibitem [{\citenamefont {{LiteBIRD Collaboration}}(2022)}]{LiteBIRD:2022cnt}%
  \BibitemOpen
  \bibfield  {author} {\bibinfo {author} {\bibnamefont {{LiteBIRD
  Collaboration}}},\ }\bibfield  {title} {\bibinfo {title} {{Probing Cosmic
  Inflation with the LiteBIRD Cosmic Microwave Background Polarization
  Survey}},\ }\href@noop {} {\bibfield  {journal} {\bibinfo  {journal} {arXiv
  e-prints}\ ,\ \bibinfo {eid} {arXiv:2202.02773}} (\bibinfo {year} {2022})},\
  \Eprint {https://arxiv.org/abs/2202.02773} {arXiv:2202.02773 [astro-ph.IM]}
  \BibitemShut {NoStop}%
\bibitem [{Note2()}]{Note2}%
  \BibitemOpen
  \bibinfo {note} {In a similar manner, there is a dependence on the
  pivot-scale chosen in fitting the isocurvature spectral index, as
  demonstrated in the $n_\protect \mathrm {iso}$ panel of figure 21 of
  Ref.~\cite {Kurki-Suonio:2004bou}.}\BibitemShut {Stop}%
\bibitem [{\citenamefont {Kurki-Suonio}\ \emph {et~al.}(2005)\citenamefont
  {Kurki-Suonio}, \citenamefont {Muhonen},\ and\ \citenamefont
  {Valiviita}}]{Kurki-Suonio:2004bou}%
  \BibitemOpen
  \bibfield  {author} {\bibinfo {author} {\bibfnamefont {H.}~\bibnamefont
  {Kurki-Suonio}}, \bibinfo {author} {\bibfnamefont {V.}~\bibnamefont
  {Muhonen}},\ and\ \bibinfo {author} {\bibfnamefont {J.}~\bibnamefont
  {Valiviita}},\ }\bibfield  {title} {\bibinfo {title} {{Correlated primordial
  perturbations in light of CMB and LSS data}},\ }\href
  {https://doi.org/10.1103/PhysRevD.71.063005} {\bibfield  {journal} {\bibinfo
  {journal} {Phys. Rev. D}\ }\textbf {\bibinfo {volume} {71}},\ \bibinfo
  {pages} {063005} (\bibinfo {year} {2005})},\ \Eprint
  {https://arxiv.org/abs/astro-ph/0412439} {arXiv:astro-ph/0412439}
  \BibitemShut {NoStop}%
\bibitem [{\citenamefont {Keskitalo}\ \emph {et~al.}(2007)\citenamefont
  {Keskitalo}, \citenamefont {Kurki-Suonio}, \citenamefont {Muhonen},\ and\
  \citenamefont {Valiviita}}]{Keskitalo:2006qv}%
  \BibitemOpen
  \bibfield  {author} {\bibinfo {author} {\bibfnamefont {R.}~\bibnamefont
  {Keskitalo}}, \bibinfo {author} {\bibfnamefont {H.}~\bibnamefont
  {Kurki-Suonio}}, \bibinfo {author} {\bibfnamefont {V.}~\bibnamefont
  {Muhonen}},\ and\ \bibinfo {author} {\bibfnamefont {J.}~\bibnamefont
  {Valiviita}},\ }\bibfield  {title} {\bibinfo {title} {{Hints of Isocurvature
  Perturbations in the Cosmic Microwave Background}},\ }\href
  {https://doi.org/10.1088/1475-7516/2007/09/008} {\bibfield  {journal}
  {\bibinfo  {journal} {JCAP}\ }\textbf {\bibinfo {volume} {09}},\ \bibinfo
  {pages} {008}},\ \Eprint {https://arxiv.org/abs/astro-ph/0611917}
  {arXiv:astro-ph/0611917} \BibitemShut {NoStop}%
\bibitem [{Note3()}]{Note3}%
  \BibitemOpen
  \bibinfo {note} {Equation~(\ref {matInter}) specifies a straight line in the
  $(\ln k,\protect \, \ln {\protect \cal P}_\protect \mathrm {t})$ plane. This
  line goes through points $(\ln k_1,\protect \, \ln {\protect \cal
  P}_{\protect \mathrm {t}1})$ and $(\ln k_2,\protect \, \ln {\protect \cal
  P}_{\protect \mathrm {t}2})$, where ${\protect \cal P}_{\protect \mathrm
  {t}1} = {\protect \cal P}_{\protect \mathrm {t}}(k_1) = r_1 {\protect \cal
  P}_{\protect \mathrm {\protect \cal R}} (k_1)$ and ${\protect \cal
  P}_{\protect \mathrm {t}2} = {\protect \cal P}_{\protect \mathrm {t}}(k_2) =
  r_2 {\protect \cal P}_{\protect \mathrm {\protect \cal R}}
  (k_2)$.}\BibitemShut {Stop}%
\bibitem [{\citenamefont {Aghanim}\ \emph {et~al.}(2020)\citenamefont {Aghanim}
  \emph {et~al.}}]{Planck:2019nip}%
  \BibitemOpen
  \bibfield  {author} {\bibinfo {author} {\bibfnamefont {N.}~\bibnamefont
  {Aghanim}} \emph {et~al.} (\bibinfo {collaboration} {Planck}),\ }\bibfield
  {title} {\bibinfo {title} {{Planck 2018 results. V. CMB power spectra and
  likelihoods}},\ }\href {https://doi.org/10.1051/0004-6361/201936386}
  {\bibfield  {journal} {\bibinfo  {journal} {Astron. Astrophys.}\ }\textbf
  {\bibinfo {volume} {641}},\ \bibinfo {pages} {A5} (\bibinfo {year} {2020})},\
  \Eprint {https://arxiv.org/abs/1907.12875} {arXiv:1907.12875 [astro-ph.CO]}
  \BibitemShut {NoStop}%
\bibitem [{\citenamefont {Lewis}\ and\ \citenamefont
  {Bridle}(2002)}]{Lewis:2002ah}%
  \BibitemOpen
  \bibfield  {author} {\bibinfo {author} {\bibfnamefont {A.}~\bibnamefont
  {Lewis}}\ and\ \bibinfo {author} {\bibfnamefont {S.}~\bibnamefont {Bridle}},\
  }\bibfield  {title} {\bibinfo {title} {{Cosmological parameters from CMB and
  other data: A Monte Carlo approach}},\ }\href
  {https://doi.org/10.1103/PhysRevD.66.103511} {\bibfield  {journal} {\bibinfo
  {journal} {\prd}\ }\textbf {\bibinfo {volume} {66}},\ \bibinfo {pages}
  {103511} (\bibinfo {year} {2002})},\ \Eprint
  {https://arxiv.org/abs/astro-ph/0205436} {arXiv:astro-ph/0205436 [astro-ph]}
  \BibitemShut {NoStop}%
\bibitem [{\citenamefont {Lewis}(2013)}]{Lewis:2013hha}%
  \BibitemOpen
  \bibfield  {author} {\bibinfo {author} {\bibfnamefont {A.}~\bibnamefont
  {Lewis}},\ }\bibfield  {title} {\bibinfo {title} {{Efficient sampling of fast
  and slow cosmological parameters}},\ }\href
  {https://doi.org/10.1103/PhysRevD.87.103529} {\bibfield  {journal} {\bibinfo
  {journal} {\prd}\ }\textbf {\bibinfo {volume} {87}},\ \bibinfo {pages}
  {103529} (\bibinfo {year} {2013})},\ \Eprint
  {https://arxiv.org/abs/1304.4473} {arXiv:1304.4473 [astro-ph.CO]}
  \BibitemShut {NoStop}%
\bibitem [{\citenamefont {Lewis}\ \emph {et~al.}(2000)\citenamefont {Lewis},
  \citenamefont {Challinor},\ and\ \citenamefont {Lasenby}}]{Lewis:1999bs}%
  \BibitemOpen
  \bibfield  {author} {\bibinfo {author} {\bibfnamefont {A.}~\bibnamefont
  {Lewis}}, \bibinfo {author} {\bibfnamefont {A.}~\bibnamefont {Challinor}},\
  and\ \bibinfo {author} {\bibfnamefont {A.}~\bibnamefont {Lasenby}},\
  }\bibfield  {title} {\bibinfo {title} {{Efficient computation of CMB
  anisotropies in closed FRW models}},\ }\href {https://doi.org/10.1086/309179}
  {\bibfield  {journal} {\bibinfo  {journal} {\apj}\ }\textbf {\bibinfo
  {volume} {538}},\ \bibinfo {pages} {473} (\bibinfo {year} {2000})},\ \Eprint
  {https://arxiv.org/abs/astro-ph/9911177} {arXiv:astro-ph/9911177 [astro-ph]}
  \BibitemShut {NoStop}%
\bibitem [{\citenamefont {Howlett}\ \emph {et~al.}(2012)\citenamefont
  {Howlett}, \citenamefont {Lewis}, \citenamefont {Hall},\ and\ \citenamefont
  {Challinor}}]{Howlett:2012mh}%
  \BibitemOpen
  \bibfield  {author} {\bibinfo {author} {\bibfnamefont {C.}~\bibnamefont
  {Howlett}}, \bibinfo {author} {\bibfnamefont {A.}~\bibnamefont {Lewis}},
  \bibinfo {author} {\bibfnamefont {A.}~\bibnamefont {Hall}},\ and\ \bibinfo
  {author} {\bibfnamefont {A.}~\bibnamefont {Challinor}},\ }\bibfield  {title}
  {\bibinfo {title} {{CMB power spectrum parameter degeneracies in the era of
  precision cosmology}},\ }\href
  {https://doi.org/10.1088/1475-7516/2012/04/027} {\bibfield  {journal}
  {\bibinfo  {journal} {JCAP}\ }\textbf {\bibinfo {volume} {1204}},\ \bibinfo
  {pages} {027}},\ \Eprint {https://arxiv.org/abs/1201.3654} {arXiv:1201.3654
  [astro-ph.CO]} \BibitemShut {NoStop}%
\bibitem [{\citenamefont {Abbott}\ \emph {et~al.}(2017)\citenamefont {Abbott}
  \emph {et~al.}}]{LIGOScientific:2016jlg}%
  \BibitemOpen
  \bibfield  {author} {\bibinfo {author} {\bibfnamefont {B.~P.}\ \bibnamefont
  {Abbott}} \emph {et~al.} (\bibinfo {collaboration} {LIGO Scientific,
  Virgo}),\ }\bibfield  {title} {\bibinfo {title} {{Upper Limits on the
  Stochastic Gravitational-Wave Background from Advanced LIGO\textquoteright{}s
  First Observing Run}},\ }\href
  {https://doi.org/10.1103/PhysRevLett.118.121101} {\bibfield  {journal}
  {\bibinfo  {journal} {Phys. Rev. Lett.}\ }\textbf {\bibinfo {volume} {118}},\
  \bibinfo {pages} {121101} (\bibinfo {year} {2017})},\ \bibinfo {note}
  {[Erratum: Phys.Rev.Lett. 119, 029901 (2017)]},\ \Eprint
  {https://arxiv.org/abs/1612.02029} {arXiv:1612.02029 [gr-qc]} \BibitemShut
  {NoStop}%
\bibitem [{\citenamefont {Valiviita}\ and\ \citenamefont
  {Giannantonio}(2009)}]{Valiviita:2009bp}%
  \BibitemOpen
  \bibfield  {author} {\bibinfo {author} {\bibfnamefont {J.}~\bibnamefont
  {Valiviita}}\ and\ \bibinfo {author} {\bibfnamefont {T.}~\bibnamefont
  {Giannantonio}},\ }\bibfield  {title} {\bibinfo {title} {{Constraints on
  primordial isocurvature perturbations and spatial curvature by Bayesian model
  selection}},\ }\href {https://doi.org/10.1103/PhysRevD.80.123516} {\bibfield
  {journal} {\bibinfo  {journal} {Phys. Rev. D}\ }\textbf {\bibinfo {volume}
  {80}},\ \bibinfo {pages} {123516} (\bibinfo {year} {2009})},\ \Eprint
  {https://arxiv.org/abs/0909.5190} {arXiv:0909.5190 [astro-ph.CO]}
  \BibitemShut {NoStop}%
\bibitem [{\citenamefont {Valiviita}\ \emph {et~al.}(2012)\citenamefont
  {Valiviita}, \citenamefont {Savelainen}, \citenamefont {Talvitie},
  \citenamefont {Kurki-Suonio},\ and\ \citenamefont
  {Rusak}}]{Valiviita:2012ub}%
  \BibitemOpen
  \bibfield  {author} {\bibinfo {author} {\bibfnamefont {J.}~\bibnamefont
  {Valiviita}}, \bibinfo {author} {\bibfnamefont {M.}~\bibnamefont
  {Savelainen}}, \bibinfo {author} {\bibfnamefont {M.}~\bibnamefont
  {Talvitie}}, \bibinfo {author} {\bibfnamefont {H.}~\bibnamefont
  {Kurki-Suonio}},\ and\ \bibinfo {author} {\bibfnamefont {S.}~\bibnamefont
  {Rusak}},\ }\bibfield  {title} {\bibinfo {title} {{Constraints on scalar and
  tensor perturbations in phenomenological and two-field inflation models:
  Bayesian evidences for primordial isocurvature and tensor modes}},\ }\href
  {https://doi.org/10.1088/0004-637X/753/2/151} {\bibfield  {journal} {\bibinfo
   {journal} {Astrophys. J.}\ }\textbf {\bibinfo {volume} {753}},\ \bibinfo
  {pages} {151} (\bibinfo {year} {2012})},\ \Eprint
  {https://arxiv.org/abs/1202.2852} {arXiv:1202.2852 [astro-ph.CO]}
  \BibitemShut {NoStop}%
\bibitem [{\citenamefont {{Galloni}}\ \emph {et~al.}(2022)\citenamefont
  {{Galloni}}, \citenamefont {{Bartolo}}, \citenamefont {{Matarrese}},
  \citenamefont {{Migliaccio}}, \citenamefont {{Ricciardone}},\ and\
  \citenamefont {{Vittorio}}}]{Galloni:2022mok}%
  \BibitemOpen
  \bibfield  {author} {\bibinfo {author} {\bibfnamefont {G.}~\bibnamefont
  {{Galloni}}}, \bibinfo {author} {\bibfnamefont {N.}~\bibnamefont
  {{Bartolo}}}, \bibinfo {author} {\bibfnamefont {S.}~\bibnamefont
  {{Matarrese}}}, \bibinfo {author} {\bibfnamefont {M.}~\bibnamefont
  {{Migliaccio}}}, \bibinfo {author} {\bibfnamefont {A.}~\bibnamefont
  {{Ricciardone}}},\ and\ \bibinfo {author} {\bibfnamefont {N.}~\bibnamefont
  {{Vittorio}}},\ }\bibfield  {title} {\bibinfo {title} {{Updated constraints
  on amplitude and tilt of the tensor primordial spectrum}},\ }\href@noop {}
  {\bibfield  {journal} {\bibinfo  {journal} {arXiv e-prints}\ ,\ \bibinfo
  {eid} {arXiv:2208.00188}} (\bibinfo {year} {2022})},\ \Eprint
  {https://arxiv.org/abs/2208.00188} {arXiv:2208.00188 [astro-ph.CO]}
  \BibitemShut {NoStop}%
\bibitem [{Note4()}]{Note4}%
  \BibitemOpen
  \bibinfo {note} {As the two-dimensional analysis of $(r_1,\protect \,r_2)$
  does not indicate any detection of a non-zero tensor contribution, i.e., the
  best fit is very near to $(0,\protect \,0)$ and $(0,\protect \,0)$ is in the
  68\% CL region, we report the conservative 95\% CL upper bounds on the
  tensor-to-scalar ratio by forcing a one-tail analysis in \protect \texttt
  {getdist}.}\BibitemShut {Stop}%
\bibitem [{\citenamefont {Ade}\ \emph {et~al.}(2019)\citenamefont {Ade} \emph
  {et~al.}}]{SimonsObservatory:2018koc}%
  \BibitemOpen
  \bibfield  {author} {\bibinfo {author} {\bibfnamefont {P.}~\bibnamefont
  {Ade}} \emph {et~al.} (\bibinfo {collaboration} {Simons Observatory}),\
  }\bibfield  {title} {\bibinfo {title} {{The Simons Observatory: Science goals
  and forecasts}},\ }\href {https://doi.org/10.1088/1475-7516/2019/02/056}
  {\bibfield  {journal} {\bibinfo  {journal} {JCAP}\ }\textbf {\bibinfo
  {volume} {02}},\ \bibinfo {pages} {056}},\ \Eprint
  {https://arxiv.org/abs/1808.07445} {arXiv:1808.07445 [astro-ph.CO]}
  \BibitemShut {NoStop}%
\bibitem [{\citenamefont {Addamo}\ \emph {et~al.}(2021)\citenamefont {Addamo}
  \emph {et~al.}}]{LSPE:2020uos}%
  \BibitemOpen
  \bibfield  {author} {\bibinfo {author} {\bibfnamefont {G.}~\bibnamefont
  {Addamo}} \emph {et~al.} (\bibinfo {collaboration} {LSPE}),\ }\bibfield
  {title} {\bibinfo {title} {{The large scale polarization explorer (LSPE) for
  CMB measurements: performance forecast}},\ }\href
  {https://doi.org/10.1088/1475-7516/2021/08/008} {\bibfield  {journal}
  {\bibinfo  {journal} {JCAP}\ }\textbf {\bibinfo {volume} {08}},\ \bibinfo
  {pages} {008}},\ \Eprint {https://arxiv.org/abs/2008.11049} {arXiv:2008.11049
  [astro-ph.IM]} \BibitemShut {NoStop}%
\bibitem [{\citenamefont {{Abazajian}}\ and\ \citenamefont
  {Others}(2016)}]{CMB-S4:2016ple}%
  \BibitemOpen
  \bibfield  {author} {\bibinfo {author} {\bibfnamefont {K.~N.}\ \bibnamefont
  {{Abazajian}}}\ and\ \bibinfo {author} {\bibnamefont {Others}},\ }\bibfield
  {title} {\bibinfo {title} {{CMB-S4 Science Book, First Edition}},\
  }\href@noop {} {\bibfield  {journal} {\bibinfo  {journal} {arXiv e-prints}\
  ,\ \bibinfo {eid} {arXiv:1610.02743}} (\bibinfo {year} {2016})},\ \Eprint
  {https://arxiv.org/abs/1610.02743} {arXiv:1610.02743 [astro-ph.CO]}
  \BibitemShut {NoStop}%
\bibitem [{\citenamefont {Baldi}\ \emph {et~al.}(2005)\citenamefont {Baldi},
  \citenamefont {Finelli},\ and\ \citenamefont {Matarrese}}]{Baldi:2005gk}%
  \BibitemOpen
  \bibfield  {author} {\bibinfo {author} {\bibfnamefont {M.}~\bibnamefont
  {Baldi}}, \bibinfo {author} {\bibfnamefont {F.}~\bibnamefont {Finelli}},\
  and\ \bibinfo {author} {\bibfnamefont {S.}~\bibnamefont {Matarrese}},\
  }\bibfield  {title} {\bibinfo {title} {{Inflation with violation of the null
  energy condition}},\ }\href {https://doi.org/10.1103/PhysRevD.72.083504}
  {\bibfield  {journal} {\bibinfo  {journal} {Phys. Rev. D}\ }\textbf {\bibinfo
  {volume} {72}},\ \bibinfo {pages} {083504} (\bibinfo {year} {2005})},\
  \Eprint {https://arxiv.org/abs/astro-ph/0505552} {arXiv:astro-ph/0505552}
  \BibitemShut {NoStop}%
\bibitem [{\citenamefont {{Kobayashi}}\ \emph {et~al.}(2011)\citenamefont
  {{Kobayashi}}, \citenamefont {{Yamaguchi}},\ and\ \citenamefont
  {{Yokoyama}}}]{2011PhRvD..83j3524K}%
  \BibitemOpen
  \bibfield  {author} {\bibinfo {author} {\bibfnamefont {T.}~\bibnamefont
  {{Kobayashi}}}, \bibinfo {author} {\bibfnamefont {M.}~\bibnamefont
  {{Yamaguchi}}},\ and\ \bibinfo {author} {\bibfnamefont {J.}~\bibnamefont
  {{Yokoyama}}},\ }\bibfield  {title} {\bibinfo {title} {{Primordial
  non-Gaussianity from G inflation}},\ }\href
  {https://doi.org/10.1103/PhysRevD.83.103524} {\bibfield  {journal} {\bibinfo
  {journal} {\prd}\ }\textbf {\bibinfo {volume} {83}},\ \bibinfo {eid} {103524}
  (\bibinfo {year} {2011})},\ \Eprint {https://arxiv.org/abs/1103.1740}
  {arXiv:1103.1740 [hep-th]} \BibitemShut {NoStop}%
\bibitem [{\citenamefont {Ade}\ \emph {et~al.}(2016{\natexlab{c}})\citenamefont
  {Ade} \emph {et~al.}}]{Planck:2015zfm}%
  \BibitemOpen
  \bibfield  {author} {\bibinfo {author} {\bibfnamefont {P.~A.~R.}\
  \bibnamefont {Ade}} \emph {et~al.} (\bibinfo {collaboration} {Planck}),\
  }\bibfield  {title} {\bibinfo {title} {{Planck 2015 results. XVII.
  Constraints on primordial non-Gaussianity}},\ }\href
  {https://doi.org/10.1051/0004-6361/201525836} {\bibfield  {journal} {\bibinfo
   {journal} {Astron. Astrophys.}\ }\textbf {\bibinfo {volume} {594}},\
  \bibinfo {pages} {A17} (\bibinfo {year} {2016}{\natexlab{c}})},\ \Eprint
  {https://arxiv.org/abs/1502.01592} {arXiv:1502.01592 [astro-ph.CO]}
  \BibitemShut {NoStop}%
\bibitem [{\citenamefont {Akrami}\ \emph
  {et~al.}(2020{\natexlab{b}})\citenamefont {Akrami} \emph
  {et~al.}}]{Planck:2019kim}%
  \BibitemOpen
  \bibfield  {author} {\bibinfo {author} {\bibfnamefont {Y.}~\bibnamefont
  {Akrami}} \emph {et~al.} (\bibinfo {collaboration} {Planck}),\ }\bibfield
  {title} {\bibinfo {title} {{Planck 2018 results. IX. Constraints on
  primordial non-Gaussianity}},\ }\href
  {https://doi.org/10.1051/0004-6361/201935891} {\bibfield  {journal} {\bibinfo
   {journal} {Astron. Astrophys.}\ }\textbf {\bibinfo {volume} {641}},\
  \bibinfo {pages} {A9} (\bibinfo {year} {2020}{\natexlab{b}})},\ \Eprint
  {https://arxiv.org/abs/1905.05697} {arXiv:1905.05697 [astro-ph.CO]}
  \BibitemShut {NoStop}%
\bibitem [{\citenamefont {Knox}\ and\ \citenamefont
  {Song}(2002)}]{Knox:2002pe}%
  \BibitemOpen
  \bibfield  {author} {\bibinfo {author} {\bibfnamefont {L.}~\bibnamefont
  {Knox}}\ and\ \bibinfo {author} {\bibfnamefont {Y.-S.}\ \bibnamefont
  {Song}},\ }\bibfield  {title} {\bibinfo {title} {{A Limit on the
  detectability of the energy scale of inflation}},\ }\href
  {https://doi.org/10.1103/PhysRevLett.89.011303} {\bibfield  {journal}
  {\bibinfo  {journal} {Phys. Rev. Lett.}\ }\textbf {\bibinfo {volume} {89}},\
  \bibinfo {pages} {011303} (\bibinfo {year} {2002})},\ \Eprint
  {https://arxiv.org/abs/astro-ph/0202286} {arXiv:astro-ph/0202286}
  \BibitemShut {NoStop}%
\bibitem [{\citenamefont {Hiramatsu}\ \emph {et~al.}(2018)\citenamefont
  {Hiramatsu}, \citenamefont {Komatsu}, \citenamefont {Hazumi},\ and\
  \citenamefont {Sasaki}}]{Hiramatsu:2018nfa}%
  \BibitemOpen
  \bibfield  {author} {\bibinfo {author} {\bibfnamefont {T.}~\bibnamefont
  {Hiramatsu}}, \bibinfo {author} {\bibfnamefont {E.}~\bibnamefont {Komatsu}},
  \bibinfo {author} {\bibfnamefont {M.}~\bibnamefont {Hazumi}},\ and\ \bibinfo
  {author} {\bibfnamefont {M.}~\bibnamefont {Sasaki}},\ }\bibfield  {title}
  {\bibinfo {title} {{Reconstruction of primordial tensor power spectra from
  B-mode polarization of the cosmic microwave background}},\ }\href
  {https://doi.org/10.1103/PhysRevD.97.123511} {\bibfield  {journal} {\bibinfo
  {journal} {Phys. Rev. D}\ }\textbf {\bibinfo {volume} {97}},\ \bibinfo
  {pages} {123511} (\bibinfo {year} {2018})},\ \Eprint
  {https://arxiv.org/abs/1803.00176} {arXiv:1803.00176 [astro-ph.CO]}
  \BibitemShut {NoStop}%
\bibitem [{\citenamefont {Campeti}\ \emph {et~al.}(2019)\citenamefont
  {Campeti}, \citenamefont {Poletti},\ and\ \citenamefont
  {Baccigalupi}}]{Campeti:2019ylm}%
  \BibitemOpen
  \bibfield  {author} {\bibinfo {author} {\bibfnamefont {P.}~\bibnamefont
  {Campeti}}, \bibinfo {author} {\bibfnamefont {D.}~\bibnamefont {Poletti}},\
  and\ \bibinfo {author} {\bibfnamefont {C.}~\bibnamefont {Baccigalupi}},\
  }\bibfield  {title} {\bibinfo {title} {{Principal component analysis of the
  primordial tensor power spectrum}},\ }\href
  {https://doi.org/10.1088/1475-7516/2019/09/055} {\bibfield  {journal}
  {\bibinfo  {journal} {JCAP}\ }\textbf {\bibinfo {volume} {09}},\ \bibinfo
  {pages} {055}},\ \Eprint {https://arxiv.org/abs/1905.08200} {arXiv:1905.08200
  [astro-ph.CO]} \BibitemShut {NoStop}%
\end{thebibliography}%

\end{document}